\documentclass[twocolumn,trackchanges]{aastex701}

\usepackage{longtable}     
\usepackage{physics}   
\usepackage{natbib}
\usepackage{tabularx}
\usepackage{makecell}
\usepackage{threeparttable}
\hypersetup{colorlinks=True, linkcolor=blue, citecolor=blue, urlcolor=blue}

\newcommand{\UA}{\affiliation{Steward Observatory, University of Arizona, 933 North Cherry Avenue, Tucson, AZ 85721-0065, USA}}
\newcommand{\GeminiNorth}{\affiliation{Gemini Observatory, 670 North A`ohoku Place, Hilo, HI 96720-2700, USA}}

\newcommand{\Monash}{\affiliation{School of Physics and Astronomy, Monash University, Clayton, Australia}}
\newcommand{\OzGrav}{\affiliation{OzGrav: The ARC Center of Excellence for Gravitational Wave Discovery, Australia}}
\newcommand{\UCSD}{\affiliation{Department of Astronomy \& Astrophysics, University of California, San Diego, 9500 Gilman Drive, MC 0424, La Jolla, CA 92093-0424, USA}}
\newcommand{\UCD}{\affiliation{Department of Physics and Astronomy, University of California, Davis, 1 Shields Avenue, Davis, CA 95616-5270, USA}}
\newcommand{\Keck}{\affiliation{W.~M.~Keck Observatory, 65-1120 M\=amalahoa Highway, Kamuela, HI 96743-8431, USA}}
\newcommand{\CfA}{\affiliation{Center for Astrophysics \textbar{} Harvard \& Smithsonian, 60 Garden Street, Cambridge, MA 02138-1516, USA}}
\newcommand{\UNC}{\affiliation{Department of Physics and Astronomy, University of North Carolina, 120 East Cameron Avenue, Chapel Hill, NC 27599, USA}}

\begin{document}

\title{SN~2023zcu: A Type IIP SN with Early Flash Features}

\author[0009-0002-2621-6611]{Monalisa Dubey}
\affiliation{Aryabhatta Research Institute of Observational Sciences, Nainital-263001, India}
\affiliation{Department of Applied Physics, Mahatma Jyotiba Phule Rohilkhand University, Bareilly-243006, India} \email[show]{monalisa@aries.res.in}

\author[0000-0003-1637-267X]{Kuntal Misra}
\affiliation{Aryabhatta Research Institute of Observational Sciences, Nainital-263001, India}\email[]{kuntal@aries.res.in}

\author[0000-0001-9210-9860]{Géza Csörnyei}
\affiliation{European Southern Observatory, Karl-Schwarzschild-Straße 2, Garching-85748 , Germany}\email[]{csornyeigeza@gmail.com}

\author[0000-0001-6191-7160]{Raya Dastidar}
\affiliation{Istituto Nazionale di Astrofisica, Osservatorio Astronomico di Brera, via E. Bianchi 46, 23807 Merate (LC), Italy}\email[]{rdastidr@gmail.com}

\author[0000-0003-4253-656X]{D. Andrew Howell}
\affiliation{Department of Physics, University of California, Santa Barbara, CA 93106-9530, USA}
\affiliation{Las Cumbres Observatory, 6740 Cortona Drive, Suite 102, Goleta, CA 93117-5575, USA}\email[]{dahowell@gmail.com}

\author[orcid=0000-0003-4102-380X, gname=David, sname= Sand]{David J. Sand}
\UA \email{dsand@arizona.edu}

\author[orcid=0000-0001-8818-0795, gname=Stefano, sname=Valenti]{Stefano Valenti}
\UCD \email{valenti@ucdavis.edu}

\author[0000-0002-2636-6508]{WeiKang Zheng}
\affiliation{Department of Astronomy, University of California, Berkeley, CA 94720-3411, USA} \email[]{weikang@berkeley.edu}

\author[0000-0003-3460-0103]{Alexei~V.~Filippenko}
\affiliation{Department of Astronomy, University of California, Berkeley, CA 94720-3411, USA} 
\affiliation{Hagler Institute for Advanced Study, Texas A\&M University, 3572 TAMU, College Station, TX 77843, USA} \email[]{afillippenko@berkeley.edu}

\author[0000-0001-8738-6011]{Saurabh Jha}
\affiliation{Rutgers, The State University of New Jersey, New Brunswick, NJ 08901, USA} \email[]{saurabh@physics.rutgers.edu}

\author[0000-0003-1546-6615]{Jesper Sollerman}
\affiliation{The Oskar Klein Centre, Department of Astronomy, AlbaNova, SE-106 91 Stockholm, Sweden} \email[]{jesper@astro.su.se}

\author[0000-0001-6272-5507]{Peter Brown}
\affiliation{George P.\ and Cynthia Woods Mitchell Institute for Fundamental Physics $\&$ Astronomy, Texas A$\&$M University, 4242 TAMU, College Station, TX 77843, USA} \email[]{pbrown801@tamu.edu}

\author[orcid=0000-0002-8297-2473, gname=Kate, sname=Alexander]{Kate D. Alexander}
\UA \email{kdalexander@arizona.edu}

\author[0000-0002-1895-6639]{Moira Andrews}
\affiliation{Las Cumbres Observatory, 6740 Cortona Drive, Suite 102, Goleta, CA 93117-5575, USA} \email[]{mandrews@lco.global}

\author[orcid=0000-0003-0123-0062, gname=Jennifer, sname=Andrews]{Jennifer Andrews}
\GeminiNorth \email{Jennifer.Andrews@noirlab.edu}

\author[]{Dre Betz}
\affiliation{Department of Astronomy, University of California, Berkeley, CA 94720-3411, USA} \email[]{andreasbetz@berkeley.edu}

\author[]{Emma Born}
\affiliation{Department of Astronomy, University of California, Berkeley, CA 94720-3411, USA} \email[]{emma.born@berkeley.edu}

\author[]{Kate Bostow}
\affiliation{Department of Astronomy, University of California, Berkeley, CA 94720-3411, USA} \email[]{katebo@berkeley.edu}

\author[0000-0002-4924-444X]{K. Azalee Bostroem}
\affiliation{Steward Observatory, University of Arizona, 933 North Cherry Avenue, Tucson, AZ 85721-0065, USA} 
\affiliation{LSST-DA Catalyst Fellow} \email[]{bostroem@arizona.edu}

\author[0000-0003-1325-6235]{Se{a'}n J. Brennan}
\affiliation{The Oskar Klein Centre, Department of Astronomy, AlbaNova, SE-106 91 Stockholm, Sweden} \email[]{sean.brennan@astro.su.se}

\author[0000-0001-5955-2502]{Thomas~G.~Brink}
\affiliation{Department of Astronomy, University of California, Berkeley, CA 94720-3411, USA} \email[]{tgbrink@berkeley.edu}

\author[orcid=0000-0003-0528-202X, gname=Collin, sname=Christy]{Collin Christy}
\UA \email{collinchristy@arizona.edu}

\author[0000-0001-9984-5131]{Elma Chuang}
\affiliation{Department of Astronomy, University of California, Berkeley, CA 94720-3411, USA} \email[]{elmachuang@berkeley.edu}

\author[orcid=0000-0002-7937-6371, gname=Yize, sname=Dong]{Yize Dong}
\CfA \email{yize.dong@cfa.harvard.edu}

\author[0000-0002-0394-6745]{Naveen Dukiya}
\affiliation{Aryabhatta Research Institute of Observational Sciences, Nainital-263001, India}
\affiliation{Department of Applied Physics, Mahatma Jyotiba Phule Rohilkhand University, Bareilly-243006, India} \email[]{ndukiya@aries.res.in}

\author[0000-0003-4914-5625]{Joseph R. Farah}
\affiliation{Department of Physics, University of California, Santa Barbara, CA 93106-9530, USA} \email[]{jfarah@lco.global}

\author[orcid=0000-0003-4537-3575, gname=Noah, sname=Franz]{Noah Franz}
\UA \email{nfranz@arizona.edu}

\author[0000-0003-0209-9246]{Estefania Padilla Gonzalez}
\affiliation{Johns Hopkins University, San Martin Dr, Baltimore, MD 21210, USA} \email[]{epadill7@jh.edu}

\author[orcid=0000-0002-6703-805X, gname=Joshua, sname=Haislip]{Joshua Haislip}
\UNC \email{jhaislip@gmail.com}

\author[orcid=0000-0003-2744-4755, gname=Emily, sname=Hoang]{Emily Hoang}
\UCD \email{emthoang@ucdavis.edu}

\author[orcid=0000-0002-0832-2974, gname=Griffin, sname=Hosseinzadeh]{Griffin Hosseinzadeh}
\UCSD \email{ghosseinzadeh@ucsd.edu}

\author[orcid=0000-0002-9454-1742, gname=Brian, sname=Hsu]{Brian Hsu}
\UA \email{bhsu@arizona.edu}

\author[]{Connor Jennings}
\affiliation{Department of Astronomy, University of California, Berkeley, CA 94720-3411, USA} \email[]{cjennings2023@berkeley.edu}

\author[orcid=0000-0003-3642-5484, gname=Vladimir, sname=Kouprianov]{Vladimir Kouprianov}
\UNC \email{v.kouprianov@gmail.com}

\author[orcid=0000-0001-9589-3793, gname=Michael, sname=Lundquist]{M.~J. Lundquist}
\Keck \email{mlundquist@keck.hawaii.edu}

\author[0000-0002-9209-2787]{Colin Macrie}
\affiliation{Rutgers, The State University of New Jersey, New Brunswick, NJ 08901, USA} \email[]{cwm78@scarletmail.rutgers.edu}

\author[0000-0003-0155-2539]{Curtis McCully}
\affiliation{Department of Physics, University of California, Santa Barbara, CA 93106-9530, USA}
\affiliation{Las Cumbres Observatory, 6740 Cortona Drive, Suite 102, Goleta, CA 93117-5575, USA} \email[]{cmccully@lco.global}

\author[]{Andrew Mchaty}
\affiliation{Department of Astronomy, University of California, Berkeley, CA 94720-3411, USA} \email[]{andrew.mchaty@berkeley.edu}

\author[orcid=0009-0008-9693-4348, gname=Darshana, sname=Mehta]{Darshana Mehta}
\UCD \email{ddmehta@ucdavis.edu}

\author[]{Katie Mora}
\affiliation{Department of Astronomy, University of California, Berkeley, CA 94720-3411, USA} \email[]{katherinemora@berkeley.edu}

\author[0000-0001-9570-0584]{Megan Newsome}
\affiliation{Department of Astronomy, The University of Texas at Austin, 2515 Speedway, Stop C1400, Austin, TX 78712, USA} \email[]{newsome.megane@gmail.com}

\author[orcid=0000-0002-0744-0047, gname=Jeniveve, sname=Pearson]{Jeniveve Pearson}
\UA \email{jenivevepearson@arizona.edu}

\author[0009-0009-7665-6827]{Neil Pichay}
\affiliation{Department of Astronomy, University of California, Berkeley, CA 94720-3411, USA} \email[]{14neil@berkeley.edu}

\author[orcid=0000-0003-4175-4960, gname=Conor, sname=Ransome]{Conor Ransome}
\UA \email{cransome@arizona.edu}

\author[orcid=0000-0002-7352-7845, gname=Aravind, sname=Ravi]{Aravind P.\ Ravi}
\UCD \email{apazhayathravi@ucdavis.edu}

\author[orcid=0000-0002-5060-3673, gname=Daniel, sname=Reichart]{Daniel E.\ Reichart}
\UNC \email{reichart@physics.unc.edu}

\author[orcid=0000-0002-7015-3446, gname=Nicol\'as, sname=Meza Retamal]{Nicol\'as Meza Retamal}
\UCD \email{nemezare@ucdavis.edu}

\author[]{Sophia Risin}
\affiliation{Department of Astronomy, University of California, Berkeley, CA 94720-3411, USA} \email[]{sbrisin@berkeley.edu}

\author[orcid=0000-0002-4022-1874, gname=Manisha, sname=Shrestha]{Manisha Shrestha}
\Monash \OzGrav \email{manisha.shrestha@monash.edu}

\author[0000-0001-7881-7748]{Ajay Kumar Singh}
\affiliation{Department of Physics, Bareilly college, Bareilly , Bareilly-243006, India} \email[]{aksnmr@gmail.com}

\author[orcid=0000-0001-5510-2424, gname=Nathan, sname=Smith]{Nathan Smith}
\UA \email{nathansmith@arizona.edu}

\author[orcid=0000-0001-8073-8731, gname=Bhagya, sname=Subrayan]{Bhagya Subrayan}
\UA \email{bsubrayan@arizona.edu}

\author[0000-0003-0794-5982]{Giacomo Terreran}
\affiliation{Adler Planetarium, 1300 S DuSable Lake Shore Dr, Chicago, IL 60605, USA} \email[]{gterreran@adlerplanetarium.org}

\author[]{William Wu}
\affiliation{Department of Astronomy, University of California, Berkeley, CA 94720-3411, USA} \email[]{wuyongxuan@berkeley.edu}

\begin{abstract}
We present a detailed photometric and spectroscopic analysis of the Type IIP supernova SN~2023zcu, which exploded in the galaxy NGC~2139 (redshift $z$ = 0.006). SN~2023zcu exhibits a well-sampled light curve covering the rise, plateau, and nebular phases. It has an optically thick phase of $100.6 \pm 0.6$ d with a magnitude drop of $\sim$1.7 mag in the {\em V} band during the transition between the plateau and the nebular phases. Weak emission features in the early-time spectra indicate a low-level interaction between circumstellar material (CSM) and the SN ejecta. The spectral evolution is well sampled and exhibits a prominent P-Cygni profile of H$\alpha$, a defining characteristic of Type IIP SNe. Signatures of metal-line formation (e.g., \ion{Fe}{2}, \ion{Ca}{2} near-infrared triplet) are also evident in the spectra as the SN evolves. Spectral modeling with the radiative-transfer code \texttt{TARDIS} during the early photospheric phase (8.7--35.5 d since explosion) yields photospheric temperatures decreasing from $\sim$9,000 to $\sim$6,000 K and expansion velocities declining from $\sim$10,000 to $\sim$5,400 km s$^{-1}$. A tailored expanding photosphere method (EPM) fit based on the \texttt{TARDIS} models provides a distance estimate of $27.8 \pm 2.0$ Mpc. Nebular-phase spectra and bolometric light-curve modeling suggest a progenitor mass in the range 12--15 M$_\odot$. This thorough analysis helps to constrain progenitor properties and explosion parameters, thereby strengthening our understanding of Type IIP SNe.

\end{abstract}



\section{Introduction} \label{sec:intro}

Type II core-collapse supernovae (CCSNe) are characterized by the presence of hydrogen (H) lines in their spectra, indicating that they originate from the explosions of massive stars \citep[$\geq$8 M$_\odot$,][]{smart_2009} that retained substantial amounts of H before the explosion. Among the various subclasses of Type II CCSNe, the Type II-plateau (IIP) are the most common \citep[57.2\%,][]{Li_loss, Graur_losssnrate_2015, Shivvers_loss}. Type IIP SNe exhibit a unique characteristic of a long-lasting optical thick phase, resulting in an approximately constant luminosity `plateau' after peak brightness in the light curve. The H recombination front receding inside the shock-heated SN ejecta powers this long-duration phase. As this front reaches the base of the hydrogen envelope, there is a drop in luminosity, and the SN enters the nebular phase, which is powered by the radioactive decay energy. From a volume and time-limited survey \citep{smart_2009}, it was suggested that the possible progenitors of Type IIP SNe are red supergiant (RSG) stars with a mass range of $\sim$8--16.5 M$_\odot$.

Type IIP SNe exhibit diversity in photometric properties. The mean value of the peak luminosity in the {\em V} band is $-16.74\pm0.04$ mag \citep{Anderson_2014} with the {\em V} band magnitude ranging from $-18.66$ \citep[SN~2021dbg;][]{2021dbg_Zhao} to $-13.77$  \citep[SN~1999br;][]{Pastorello_1999br}. Diversity in plateau duration is also observed, where the majority of Type IIP SNe show normal plateau durations of 80--120 d \citep{Valenti_2016}. A few of them exhibit shorter plateaus (e.g., SNe~1995ad, \citealt{McNaught_1995ad}; 2006Y, 2006ai, 2016egz, \citealt{Hirmatshu_shortP}; 2018gj, \citealt{Teja_2018gj}; 2020jfo, \citealt{Teja_2022, Ailawadhi_2023}; 2023ufx, \citealt{Ravi_2023ufx}), while some show a long-lasting plateau phase (e.g., SNe~2005cs, \citealt{Pastorello_2005cs_2009}; 2009ib, \citealt{2009ib_takats}; 2015ba, \citealt{Raya_2015ba}; 2018hwm, \citealt{Reguitti_2018hwm}; 2020cxd, 2021aai, \citealt{2020cxd_2021aai_Valerian}). After the plateau phase, the luminosity typically drops by 1.0--2.6 mag \citep{Valenti_2016}. However, in some cases, a larger drop is noticed ($\sim$3.8 mag in SN~2018zd; \citealp{Hiramatsu_2018zd}, and $\sim$3.83 mag in the low-luminosity SN~2005cs; \citealp{Pastorello_2005cs_2009}). During the nebular phase, the SN is powered by the radioactive decay of synthesized $^{56}$Ni, typically in the range of 0.007--0.08 M$_\odot$, with a mean value of $0.033 \pm 0.024$ M$_\odot$ \citep{Anderson_2014, Mular_2017, Anderson_2019, Rodriguez_2021}. 

During the latest evolutionary stage, massive stars ($\geq$8 M$_\odot$) undergo enhanced mass loss \citep[10$^{-2}$--10$^{-6}$ M$_\odot$ yr$^{-1}$,][]{Smith_2014}, producing circumstellar material \citep[CSM;][]{Gal_yam_2014, Morozova_2017}. The radiation from the shock breakout ionizes the CSM and produces the narrow emission lines \citep[`flash features';][]{Khazov_2016} in the early-time spectra ($t \leq 10$\,d). In $\sim$30\% of Type II SNe, these features have been observed \citep{Bruch_2021}. Some Type II SNe show prominent flash features (e.g., SNe~2013fs, \citealp{Yaron_2013fs_2017, Chugai_2013fs}; 2023ixf, \citealp{bostroem_2023ixf, Jacob_2023ixf, smith_2023ixf, hu_2023ixf}; 2024ggi, \citealp{Shrestha_2024ggi, Jacob_24ggi_2024}), whereas others exhibit subtle emission (e.g., SNe~2017gmr, \citealp{Andrews_2017gmr}; 2021yja, \citealp{Hosseinzadeh_2021yja_2022}; 2022acko, \citealp{Lin_2022acko}). Sometimes CSM interaction does not produce enough energy to create high-ionization lines (e.g., \ion{He}{2}, \ion{C}{3}, \ion{N}{3}, \ion{N}{4}); rather, it shows a `ledge-shaped' feature \citep{Dessart_2017, Dessart_2023} in the early spectra around 4500--4800~\AA, which can be a result of possible blending of weak high-ionization lines. Type IIP SNe also exhibit spectral diversity in addition to photometric variations. Their line velocities correlate with luminosity; for example, low-luminosity events exhibit narrower spectral lines, indicating lower expansion velocities \citep{Pastorello_2004}. \cite{Gutierrez_2014} found that SNe with steeper declines in the {\em V} band plateau tend to have smaller absorption-to-emission ratios in their H$\alpha$ profiles. During the nebular phase, the SN spectra can probe deep inside the ejecta owing to the low-density medium. Theoretical studies have demonstrated that the forbidden lines ([\ion{O}{1}] 6300, 6364~\AA, [\ion{Fe}{2}] 7155~\AA, and [\ion{Ni}{2}] 7378~\AA) visible in the nebular spectra can provide a direct link to the progenitor mass \citep{Woosley_2007, Jerkstrand_2012, Jerkstrand_2014}. 

In this paper, we present photometric and spectroscopic analysis of the Type IIP SN~2023zcu. Section~\ref{sec:Obs} presents the SN properties and the observations and data reduction procedures. The photometric evolution, early light-curve modeling, comparison of the colors and absolute magnitudes, and estimation of the $^{56}$Ni mass are described in Section~\ref{sec:photometric_analysis}. Spectroscopic analysis, spectral modeling, and tailored expanding photosphere method (EPM) distance estimation are presented in Section~\ref{sec:spectroscopic_analysis}. The progenitor mass, constrained through nebular spectroscopy and bolometric light-curve modeling, is detailed in Section~\ref{sec:progenitor}. Conclusions of our findings are given in Section~\ref{sec:discussion}.

\section{Observations and data reduction}
\label{sec:Obs}
\subsection{Discovery and Explosion Epoch}
\label{sec:Dis_explosion}

\begin{figure}
    \centering
    \includegraphics[width=\linewidth]{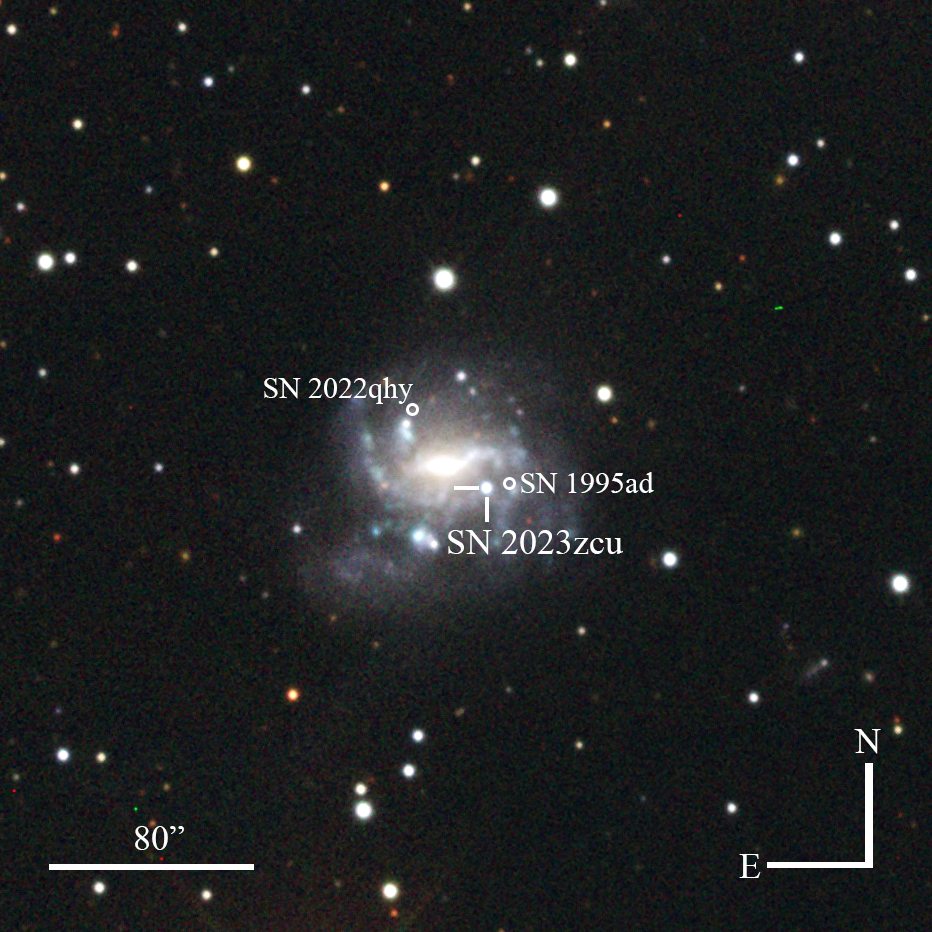}
    \caption{Color composite image of SN~2023zcu, created using {\em g}, {\em r}, and {\em i} filters, observed with the 1\,m telescope of Las Cumbres Observatory on December 18, 2023. The SN is marked in white, along with the positions of two other SNe (1995ad and 2022qhy) previously discovered in the same galaxy, NGC~2139.}
    \label{fig:2023zcu_image}
\end{figure}

SN~2023zcu (ATLAS23wvm) was discovered by the Asteroid Terrestrial-impact Last Alert System (ATLAS) survey at 19.054 (AB mag) in the {\em orange}-ATLAS filter, on UTC 2023-12-08.18 (MJD 60,286.17) with the ATLAS El Sauce telescope, Chile \citep{ATLAS_2023zcu}. Figure~\ref{fig:2023zcu_image} depicts the position of the SN (RA =  06:01:06.84, Dec. = $-$23:40:29.24, J2000.0) within the host galaxy, NGC~2139, along with the locations of two other SNe~1995ad and 2022qhy, which exploded in the same galaxy. SN~2023zcu was classified by the extended Public ESO Spectroscopic Survey of Transient Objects \citep[ePESSTO+;][]{pessto_smartt_2015} collaboration \citep{Calssification_2023zcu} based on the early spectrum acquired on UTC 2023-12-09.60 (MJD 60,287.27) by the EFOSC2 spectrograph mounted on the 3.58\,m ESO New Technology Telescope (NTT) at La Silla. The last nondetection was reported on UTC 2023-12-06.99 (MJD 60,284.99) at a limiting AB magnitude of 19.49 in the {\em orange}-ATLAS filter. The difference between the last nondetection and the first photometric detection is roughly one day, which constrains the explosion epoch to be $t_0 = \mathrm{MJD~} 60,285.59\pm0.59$. Detailed information on SN~2023zcu and its host galaxy is given in Table~\ref{tab:SN 2023zcu and Host information}. 

\begin{table}
	\centering
	\caption{Basic information on SN~2023zcu and NGC 2139.}
	\begin{tabular}{ll} 
		\hline
            \multicolumn{2}{c}{\textbf{SN~2023zcu}} \\ \hline
		Discoverer internal name & ATLAS23wvm\\
            SN type & Type IIP\\
		RA &  06:01:06.84 (J2000)\\
            Dec. & $-$23:40:29.24 (J2000)\\
	    Galactic Extinction ($A_{V,{\rm MW}}$[mag]) & 0.093\\
            Distance (Mpc)$^\ddagger$ & 27.8$\pm$2.0\\
            Discovery date (MJD) & 60,286.17\\
            Last nondetection (MJD) & 60,284.99\\
            Explosion epoch (MJD) & 60,285.59$\pm$0.59\\
            \hline\multicolumn{2}{c}{\textbf{NGC 2139}$^{\dagger}$} \\ \hline
            Galaxy type & SAB(rs)cd\\
            Major axis diameter & $147.30''$\\
            Minor axis diameter & $109.00''$\\
		Redshift & 0.006124$\pm$0.000005   \\
            Mean distance (Mpc) & 26.8$\pm$1.2\\ 
		Helio. velocity (km s$^{-1}$) & 1836$\pm$1\\
		\hline
	\end{tabular}
	\label{tab:SN 2023zcu and Host information}
        \vspace{0.2cm}
        \newline
        \noindent
        $^\ddagger${Distance of the SN is estimated using the tailored-EPM method (see Section~\ref{Tailored_EPM}).}\\
        $^\dagger${Taken from NASA/IPAC Extragalactic Database (NED).}\\
        
\end{table}

NGC~2139 is morphologically classified as an SABcd galaxy \citep{deVaucouleurs_1991}. After correcting for Virgo+GA+Shapely peculiar velocity and assuming H$_0 = 73$ km s$^{-1}$ Mpc$^{-1}$, $\Omega_{\mathrm{matter}}=0.27$, and $\Omega_{\mathrm{vacuum}}=0.73$, the host redshift is taken to be $z= 0.006124\pm0.000005$ from NED\footnote{\url{https://ned.ipac.caltech.edu/byname?objname=NGC+2139}} with a corresponding mean distance of 26.8$\pm$1.2 Mpc, which is similar to the distance estimated from tailored-EMP method (see Section~\ref{Tailored_EPM}). We adopt the tailored EPM distance for the analysis throughout the paper. The foreground Galactic reddening {\textit {E(B-V)}} = 0.03 mag is measured using the dust maps \citep{Schlafly_2011} and applying the extinction law \citep{Cardelli_1989} with $R_V = 3.1$. There is no \ion{Na}{1~D} absorption line detected in the spectra at the redshift of the host galaxy. Therefore, no contribution from the host galaxy is included in the reddening.

\subsection{Data}
\label{sec:data}

Soon after the SN discovery, photometric follow-up observations were initiated with the 1\,m telescopes of the Las Cumbres Observatory (LCO) telescope network \citep{Brown_2013} under the  Global Supernova Project (GSP) collaboration in {\em UBgVri} bands. The image pre-processing was done using the \texttt{BANZAI} pipeline \citep{McCully_2018}. The host-galaxy templates taken on UTC 2025-03-30.04 were subtracted from the science frames. Point-spread-function (PSF) photometry on the difference images was done using \texttt{lcogtsnpipe}\footnote{\url{https://github.com/LCOGT/lcogtsnpipe/}} \citep{Valenti_2016}.

Photometric observations were also carried out with 0.76\,m Katzman Automatic Imaging Telescope \citep[KAIT;] []{Filippenko_2001} and the 1\,m Nickel telescope of Lick Observatory in the {\em BVRI} bands. An automated image-reduction pipeline \citep{Ganeshalingam_2010, Stahl_2019} was used to perform the image pre-processing, while PSF photometry was done using \texttt{DAOPHOT} \citep{Stetson_1987} to calculate the instrumental magnitudes. The local standard stars from the PanSTARRS1 survey \citep{Schlafly_2012} were used for calibration. The SN was also monitored with the DLT40 {\em OPEN} filter mounted on the PROMPT telescope, and the photometric magnitudes were calibrated to the {\em r} band \citep{Tartaglia18}.

The SN was also observed with the Ultra-Violet/Optical Telescope \citep[UVOT;][]{Roming_UVOT} on \textit{The Neil Gehrels Swift Observatory} \citep[hereafter \textit{Swift};][]{Gehrels_swift}. The observations were done in UV ({\em uvw2, uvm2, uvw1}) and optical ({\em ubv}) filters. Data reduction was performed using the pipeline described in \cite{Brown_2009}.

The photometric data obtained from all telescopes are available in machine-readable format.

Low to medium-resolution spectroscopic observations, from $\sim$ 1 to 392 d since the explosion, were carried out using various telescopes and instruments listed in the log of spectroscopic observations (Table~\ref{spec_log}). Standard data reduction techniques and pipelines were adopted to extract the one-dimensional spectra. All spectra are scaled to the {\em UBgVri} photometric magnitudes using the \texttt{lightcurve-fitting}\footnote{\url{https://github.com/griffin-h/lightcurve_fitting}} \texttt{Python} package \citep{Griffin_2022}, and corrected for the redshift of the host galaxy. 

The spectra will be made publicly available in WiseRep\footnote{\url{https://www.wiserep.org/}}.

\begin{table}
\centering
\caption{Log of spectroscopic observations.}
\label{spec_log}

\resizebox{\columnwidth}{!}{%
  \begin{tabular}{cccc} \hline
  Phase$^\dagger$ &Telescope$+$Instrument & Resolution & Wavelength Range  \\
  (days) &  & ($\lambda/\Delta\lambda$) & (\AA) \\
 \hline

 1.19 & ESO-NTT$+$EFOSC & 100-2000 & 3500–10000\\
 1.28 & FTN$+$FLOYDS & 400-700 & 3300-10180\\
 1.85 & Gemini North$+$GMOS-N & 1688 & 3916-7069\\
 2.35 & FTN$+$FLOYDS & 400-700 & 3300-10180\\
 2.38 & FTN$+$FLOYDS & 400-700 & 3300-10180\\
 3.27 & SALT$+$RSS & 600-2200 & 3533-7449\\
 3.48 & NOT$+$ALFOSC & 710 & 3200-9600\\
 3.61 & ESO-VLT-UT1$+$FORS2 & 260-1600 & 3300–11000\\
 4.74 & Shane$+$Kast & 700–1500 & 3800–10000\\
 7.49 & FTS$+$FLOYDS & 400-700 & 3300-10180\\
 7.69 & MMT$+$Binospec &  1340 & 3850-9150\\
 7.71 & Bok$+$B$\&$C & 550-1100 & 4400-8700\\
 8.68 & ESO-VLT-UT1$+$FORS2 & 260-1600 & 3300–11000\\
 20.40 & FTS$+$FLOYDS & 400-700 & 3300-10180\\
 23.51 & SOAR$+$GHTS RED & 307-540 & 3000-7050\\
 26.47 & SOAR$+$GHTS RED & 307-540 & 3000-8800\\
 28.36 & FTN$+$FLOYDS & 400-700 & 3300-10180\\ 
 31.51 & SOAR$+$GHTS BLUE & 307-540 & 3000-7050\\
 35.51 & FTS$+$FLOYDS & 400-700 & 3300-10180\\
 35.69 & Shane$+$Kast & 700–1500 & 3800–10000\\
 36.53 & SOAR$+$GHTS BLUE & 307-540 & 3000-7050\\
 37.59 & MMT$+$Binospec &  1340 & 3850-9150\\
 41.34 & FTS$+$FLOYDS & 400-700 & 3300-10180\\
 46.65 & SOAR$+$GHTS RED & 307-540 & 3000-7050\\
 47.15 & FTN$+$FLOYDS & 400-700 & 3300-10180\\
 51.54 & SOAR$+$GHTS RED & 307-540 & 3000-7050\\
 55.22 & FTN$+$FLOYDS & 400-700 & 3300-10180\\
 63.38 & FTS$+$FLOYDS & 400-700 & 3300-10180\\
 66.44 & FTS$+$FLOYDS & 400-700 & 3300-10180\\
 68.51 & Bok$+$B$\&$C & 550-1100 & 4400-8700\\
 74.16 & FTN$+$FLOYDS & 400-700 & 3300-10180\\
 85.55 & MMT$+$Binospec &  1340 & 3850-9150\\
 87.39 & FTS$+$FLOYDS & 400-700 & 3300-10180\\
 95.41 & FTS$+$FLOYDS & 400-700 & 3300-10180\\
 97.32 & FTS$+$FLOYDS & 400-700 & 3300-10180\\
 101.14 & FTN$+$FLOYDS & 400-700 & 3300-10180\\
 107.14 & FTN$+$FLOYDS & 400-700 & 3300-10180\\
 109.37 & FTS$+$FLOYDS & 400-700 & 3300-10180\\
 126.33 & FTS$+$FLOYDS & 400-700 & 3300-10180\\
 141.27 & FTS$+$FLOYDS & 400-700 & 3300-10180\\
 308.00 & Keck I$+$LRIS & 1200-2200 & 3200-10200\\
 362.40 & MMT$+$Binospec &  1340 & 3850-9150\\
 366.40 & MMT$+$Binospec &  1340 & 3850-9150\\
 391.70 & Keck I$+$LRIS & 1200-2200 & 3200-10200\\ 
 \hline
 \end{tabular}%
 }

 $^\dagger$Phase relative to the explosion epoch (MJD = 60285.59). 

\end{table}

\section{Photometric Analysis}
\label{sec:photometric_analysis}

\subsection{Early Light Curve Modeling}
\label{sec:early_LC_modeling}

\begin{figure}
    \centering
    \includegraphics[width=\columnwidth]{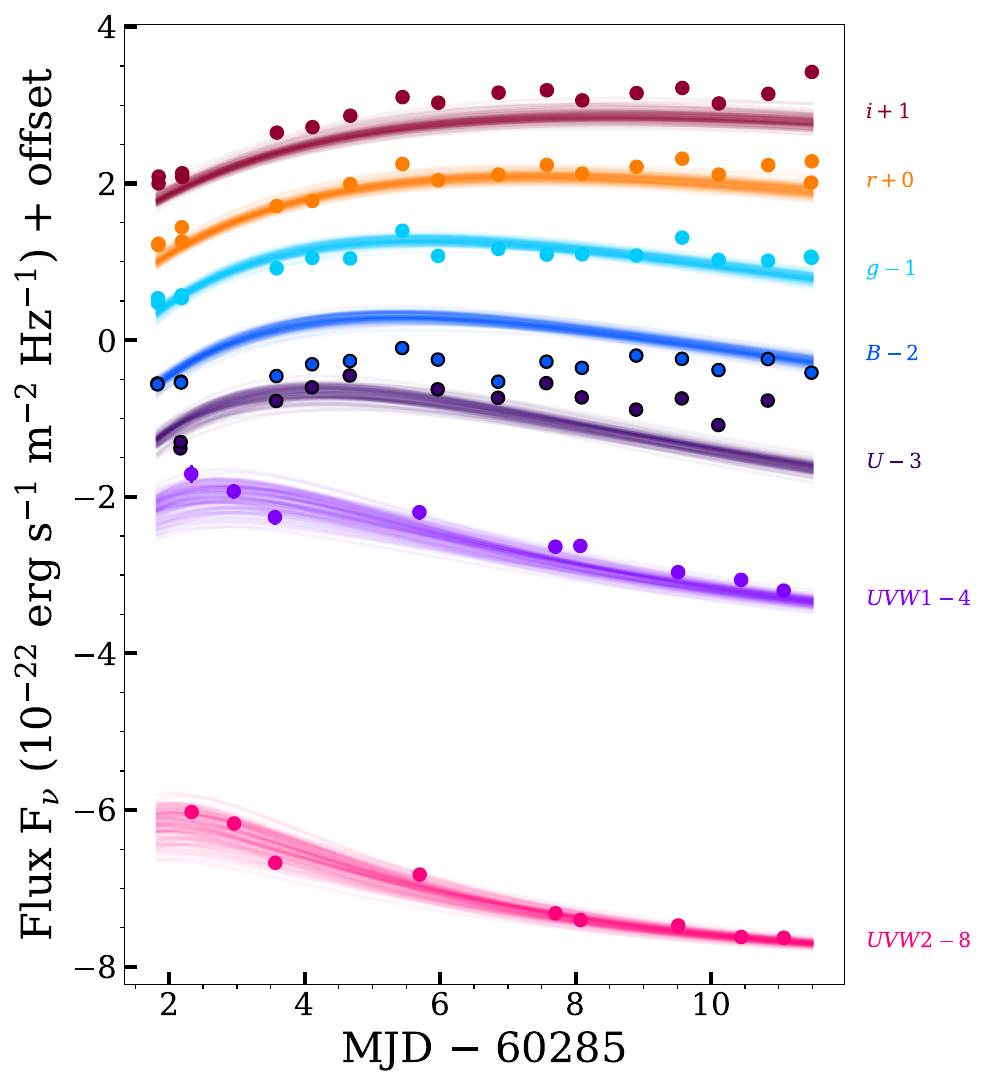}
    \caption{The early-time light curve modeling is performed using the shock-cooling model proposed by \citet{Sapir_Waxman_2017}, implemented in the \texttt{Python} code described in \citet{hosseinzadeh_2020}. The 100 best fits generated in this model using MCMC are presented along with the observed light curves.}
    \label{fig:lc_fit_UBgri}
\end{figure}

In the absence of strong CSM interaction, the early-time light curve of a SN is primarily powered by the shock-cooling mechanism, depending upon the properties of the progenitor and the explosion parameters. To obtain these parameters, the early-time UV and optical light curves of SN~2023zcu up to MJD 60297.0 (11.41 d after the explosion) are fitted using the \texttt{Python}-based light-curve fitting routine by \cite{hosseinzadeh_2020}, developed using the model proposed by \cite{Sapir_Waxman_2017}. The model gives the following physical parameters: shock velocity ($v_{s*}$), envelope mass ($M_{\rm env}$), ejecta mass ($M$), numerical factor of order unity ($f_{\rho}$), radius ($R$), including the distance ($D$) and reddening, $E(B-V)$, as free parameters. Multiplication of $f_{\rho}$ and $M$ is presented as a single parameter, which is also highly degenerate. The additional intrinsic scatter term ($\sigma$) is also included with a uniform prior to account for extra sources of uncertainty, such as photometric calibration errors, deviations from a pure blackbody spectrum, or minor variations from the model. This term is incorporated into the reported parameter uncertainties. To cover a wide wavelength range from UV to optical, we select the two \textit{Swift}/UVOT bands ({\em uvw1, uvw2}) and the {\em UBgri} optical bands for the modeling. A Markov Chain Monte Carlo (MCMC) routine and a flat prior for all parameters are used, except for $D$ and $E(B-V)$, for which Gaussian priors are used. The 100 best-fit model light curves, along with the corresponding observed light curves, are shown in Figure~\ref{fig:lc_fit_UBgri}. The parameters, prior ranges, and best-fit values are summarized in Table~\ref{tab:shock_cooling}. The rise and peak in each band are well reproduced, except the {\em B} band. The model estimates a best-fit $E(B-V) = {0.05}^{+{0.02}}_{-{0.01}}$ mag, similar to the $E(B-V)$ mentioned in Section~\ref{sec:Dis_explosion}. This model is sensitive to $R$, which depends on $E(B-V)$ \citep{Hosseinzadeh_2021yja_2022}. The best-fit value of $R$ is found to be $8.3^{+0.7}_{-0.8} \times 10^{13}$ cm ($\sim$1200$\pm$100 R$_\odot$), which is within the typical radii of RSG stars \citep[100--1500 R$_\odot$,][]{Levesque_2017}, but much higher than the radius estimated from the semi-analytical modeling presented in Section~\ref{sec:bol_LC_modeling}. The distance is evaluated as 26$^{+1}_{-2}$ Mpc, consistent with the value obtained from the tailored-EPM method (see Section~\ref{Tailored_EPM}). The explosion epoch is MJD $60285.60_{-0.05}^{+0.06}$, which is similar to the explosion epoch estimated in Section~\ref{sec:Dis_explosion}. Overall, the modeling yields estimates of distance, extinction, and explosion epoch consistent with independent determinations within their respective uncertainties, while the progenitor radius estimated from this modeling shows a discrepancy compared to that derived from the semi-analytical modeling.

\begin{table}
\begin{center}
\begingroup
\footnotesize
\caption{Shock-cooling parameters of SN~2023zcu.}
\label{tab:shock_cooling}

\begin{tabularx}{\columnwidth}{Xll}

    \hline
    \textbf{Parameter} & \textbf{Prior Range} & \textbf{Best-fit Value} \\
    \hline
    Shock Velocity [$v_{s*}$ ($10^{8.5}$ cm/s)]  & 0.0--1.0 & $0.69^{+0.05}_{-0.06}$  \\
    Envelope Mass [$M_{\rm env}$ (M$_\odot$)]  &  0.0--1.0 & $0.40\pm0.04$  \\
    Ejecta Mass $\times$ factor [$f_\rho M$  (M$_\odot$)] &  10.0--100.0 & $50\pm10$   \\
    Progenitor Radius [$R$ ($10^{13}$ cm)] &  0.0--10.0 & $8.3^{+0.7}_{-0.8}$  \\
    Distance [$d_L$ (Mpc)]  &  20--30 & $26^{+1}_{-2}$   \\
    Extinction [\textit{$E(B-V)$} (mag)]  &  0.0--0.2 & $0.05^{+0.02}_{-0.01}$ \\
    Explosion Time [$t_0$ (MJD)] &  60285.5--60287 & $60285.60^{+0.06}_{-0.05}$ \\
    Intrinsic scatter [$\sigma$ (Dimensionless)] &  0--10 & $2.0^{+0.2}_{-0.3}$ \\
    \hline
\end{tabularx}

\endgroup
\end{center}
\end{table}

\subsection{Estimation of Observational Parameters}
\label{sec:obs-params}

\begin{figure*}
    \centering
    \includegraphics[width=\linewidth]{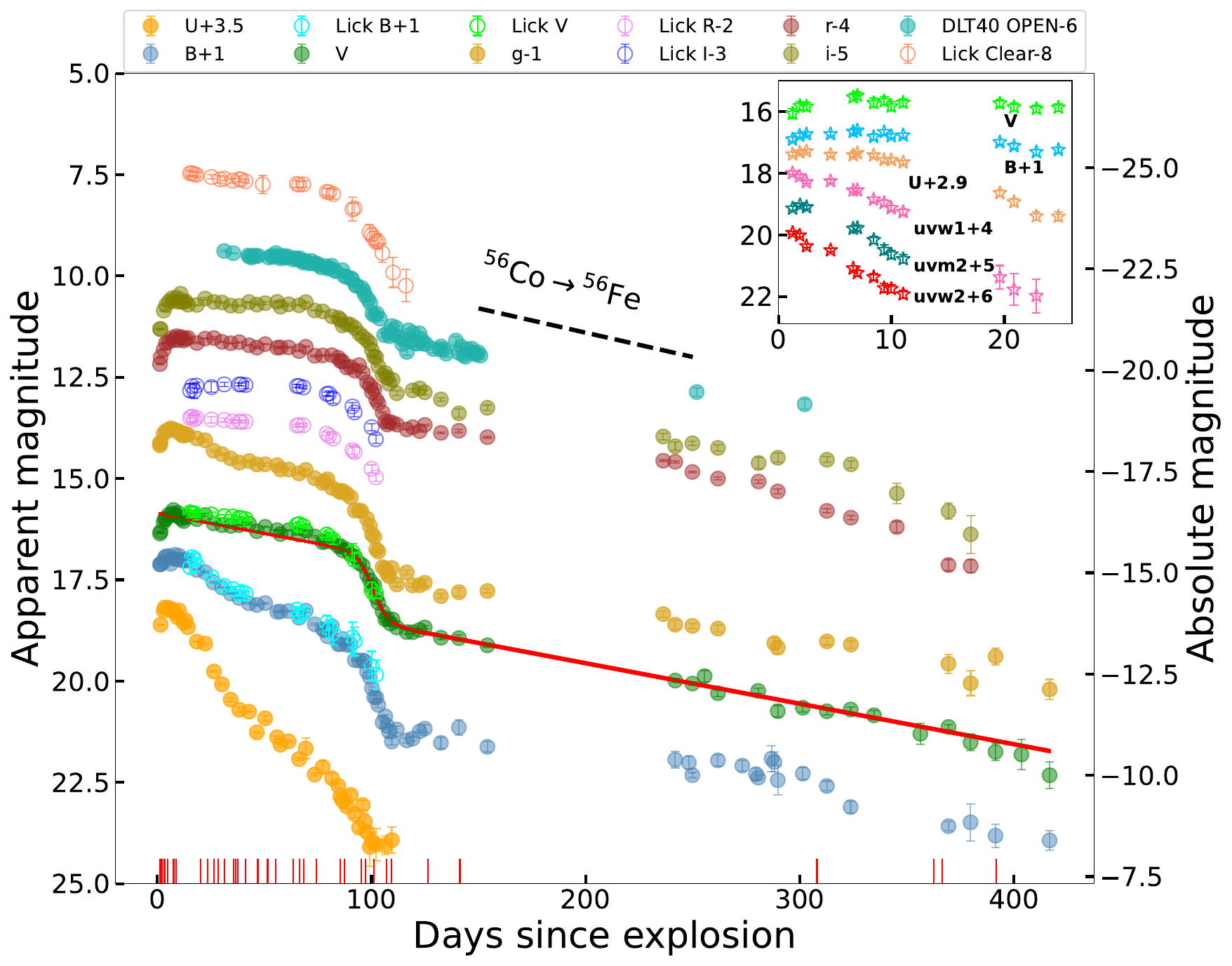}
    \caption{UV and optical light curves in apparent and absolute magnitude scales of SN~2023zcu. The \textit{Swift}/UVOT light curves are shown in the inset. The photometric data span from 1.25 to $\sim$416 d, covering the different phases of the SN. The spectral epochs are denoted with red lines at the bottom. The 100 best-fit model light curves derived from the analytical model \citep{Valenti_2016} in the {\em V} band using MCMC, are shown as red lines. (Data used to create this figure is available.)}
    \label{fig:light_curve}
\end{figure*}

Multiband optical light curves in {\em UBgVriRI}, {\em Clear}, and DLT40 {\em Open} filters, spanning from 1.25 to $\sim$416 d (in the observer frame of reference) since the explosion, are presented in Figure~\ref{fig:light_curve}. The \textit{Swift}/UVOT light curves in UV ({\em uvw2, uvm2, uvw1}), and optical ({\em ubv}) filters are shown in the inset of Figure~\ref{fig:light_curve}. Photometric data cover the rising phase, the optically thick photospheric phase (`plateau'), the drop from the plateau (`transition'), and finally the radioactive decay phase (`nebular'). The peak absolute magnitude of $-16.54\pm0.01$ is attained in the {\em V} band at 7.51 d. The bluer wavelength bands reach their peak magnitude a few days earlier than the redder ones. After the rise, the light curve initially declines, followed by the nebular phase. The decay in the plateau phase is estimated to be $s = 0.84 \pm 0.01$ mag (100 d)$^{-1}$ in the {\em V} band. The light curves in the bluer bands decline more steeply than those in the redder bands. A drop in magnitude occurs after the plateau phase, marking the `transition phase' as the recombination front reaches the base of the hydrogen envelope. This steep drop is followed by the nebular phase. The decline rate during the nebular phase in {\em V} band is denoted by $s_3$ (following the convention in \citealt{Anderson_2014}). The values of the slopes at different phases of the SN evolution, peak magnitudes, and absolute magnitudes at 50 d post-explosion in the multiband light curves are summarized in Table~\ref{tab:multiband_info}.

The analytical expression (equation~\ref{eu_Valenti}) by \cite{Valenti_2016} is fit to the {\em V} band light curve,

\begin{equation} 
y(t)= \frac{-a_0}{1+e^\frac{t-t_{\rm PT}}{w_0}} + p_0\times t + m_0\, .
\label{eu_Valenti}
\end{equation}

\noindent
Here, the depth and duration of the drop from the plateau to the nebular phase are denoted as $a_0$ and $6 \times w_0$ respectively, $t$ is the epoch of explosion, $t_{\rm PT}$ is the plateau duration in days, $p_0$ constrains the slope before and after the drop, and $m_0$ is a constant. The first term represents the Fermi-Dirac function, which describes the phase transition between the plateau and the nebular phases. To replicate the slope of the light curve before and after the drop, the second component adds the slope to the Fermi-Dirac function. The MCMC routine \texttt{emcee}\footnote{\url{https://emcee.readthedocs.io/en/stable/}} \citep{emcee_Foreman} is used to fit the above analytical function to the {\em V} band light curve. The best-fit observational parameters are estimated as $t_{\rm PT}$ = 100.57$\pm$0.59 d, $a_0$ = 1.70$\pm$0.01 mag, $w_0$ = 3.44$\pm$0.08 d, and $p_0$ = 1.00$\pm$0.01 mag (100 d)$^{-1}$. The 100 best-fit light curves obtained from the analytical function to the {\em V} band light curve are shown in Figure~\ref{fig:light_curve} with red solid lines. 

\begin{figure*}
    \centering
    \includegraphics[width=\textwidth]{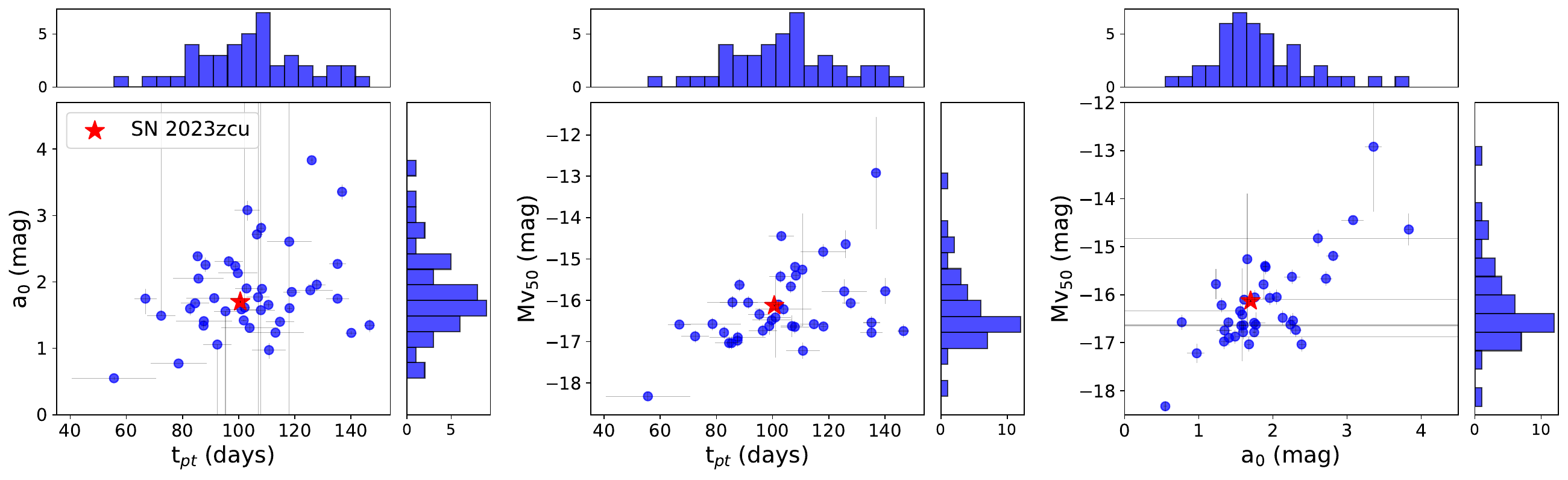}
    \caption{Correlations between $a_0$, $t_{\rm PT}$ and $M_{V50}$. The red star represents SN~2023zcu, while the blue circles denote the SNe in the samples of \citet{Anderson_2014} and \citet{Valenti_2016}.}
    \label{fig:phot_correlation}
\end{figure*}

We compare the {\em V} band absolute magnitude at 50 d post explosion ($M_{V50}$), $t_{\rm PT}$, and $a_0$ of SN~2023zcu with other Type II SNe from  \cite{Anderson_2014} and \cite{Valenti_2016}. The red star in Figure~\ref{fig:phot_correlation} represents the location of SN~2023zcu, while the blue circles correspond to the other SNe in the sample. The parameter estimates of SN~2023zcu are in agreement with the sample averages.

\begin{table*}
    \centering
    \caption{Properties of the multiband optical light curves.}
    \label{tab:multiband_info}
    \resizebox{\textwidth}{!}{%
    \begin{tabular}{lcccccccc}
        \hline
        Paramater & {\em U} & {\em B} & {\em g} & {\em V} & {\em r}  & {\em i} & {\em Clear} & {\em Open} \\
        \hline
        Peak (mag) & $-$17.64$\pm$0.01 & $-$16.43$\pm$0.01 &  $-$16.78$\pm$0.02 & $-$16.54$\pm$0.01 & $-$16.82$\pm$0.01 & $-$16.85$\pm$0.01 & - & - \\
        Rise Time (days) & 4.85 & 5.38 & 6.27 & 7.51 & 8.98 & 10.90 & -- & -- \\
        $s$ (mag (100 d)$^{-1}$) &  6.57$\pm$0.02 & 2.74$\pm$0.01 & 1.84$\pm$0.01 & 0.84$\pm$0.01 & 0.59$\pm$0.01 & 0.24$\pm$0.01 & 0.59$\pm$0.08 & 0.59$\pm$0.02 \\
        $s_3$ (mag (100 d)$^{-1}$) & - & 0.69$\pm$0.02 & 0.54$\pm$0.01 & 0.99$\pm$0.01 & 0.89$\pm$0.01 & 0.98$\pm$0.02 & - & 1.01$\pm$0.04\\
        Transition slope (mag (100 d)$^{-1}$) & - & 9.55$\pm$0.13 & 10.59$\pm$0.13 &8.18$\pm$0.09 & 7.55$\pm$0.08 &  5.96$\pm$0.05 & 5.91$\pm$0.48 & 4.45$\pm$0.03 \\
        $M_{50}$ (mag) & $-$14.95$\pm$0.01 & $-$15.28$\pm$0.01 &  $-$15.89$\pm$0.01 & $-$16.13$\pm$0.02 & $-$16.62$\pm$0.01 & $-$16.63$\pm$0.01 & $-16.55\pm0.22$ & $-16.47\pm0.02$ \\
        \hline
    \end{tabular}%
    }
\end{table*}

\subsection{Comparison of Photometric Characteristics}
\label{sec:phtotmetric_comparision}

\begin{figure}
    \centering
    \includegraphics[width=\linewidth]{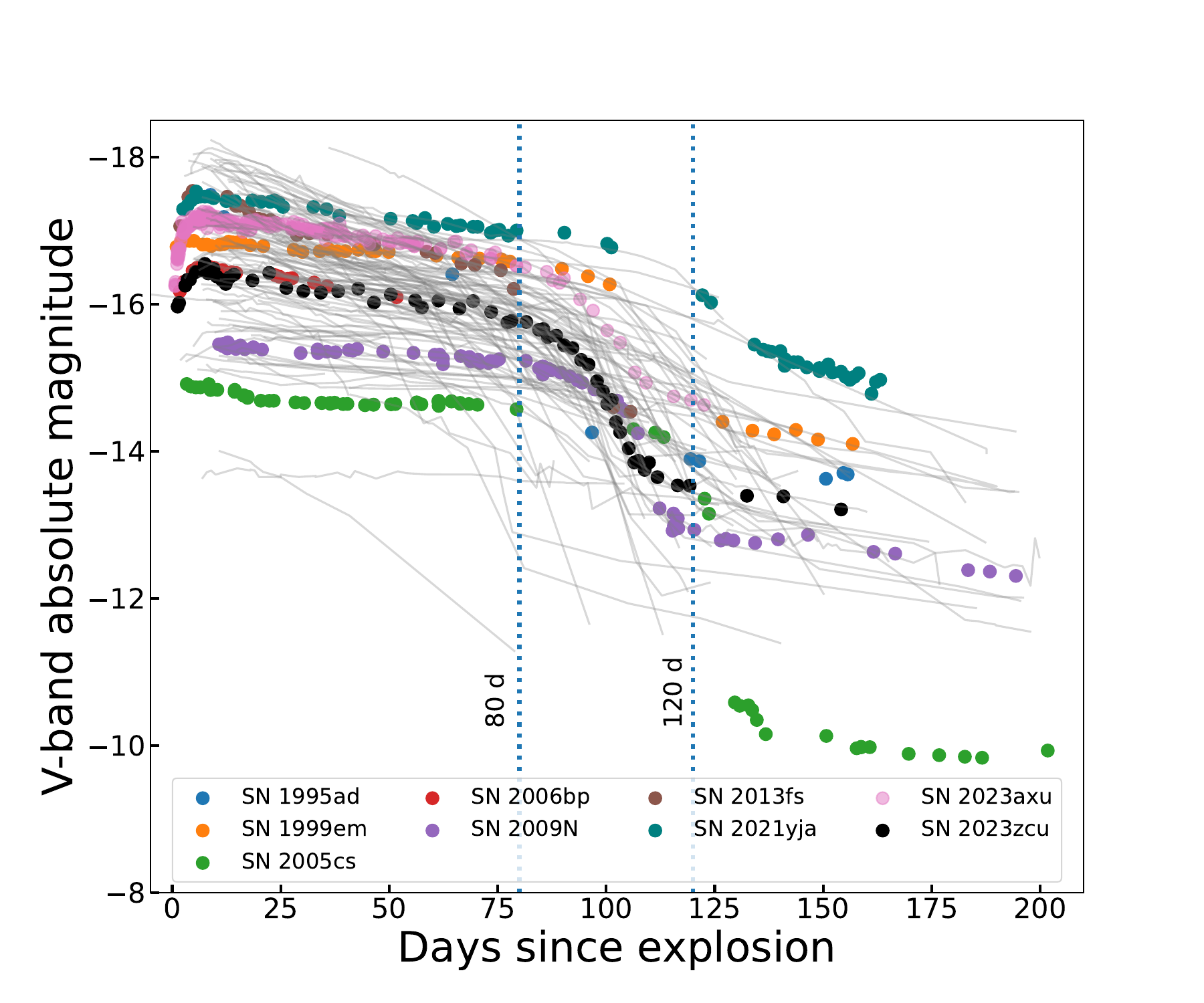}\\
    \includegraphics[width=\linewidth]{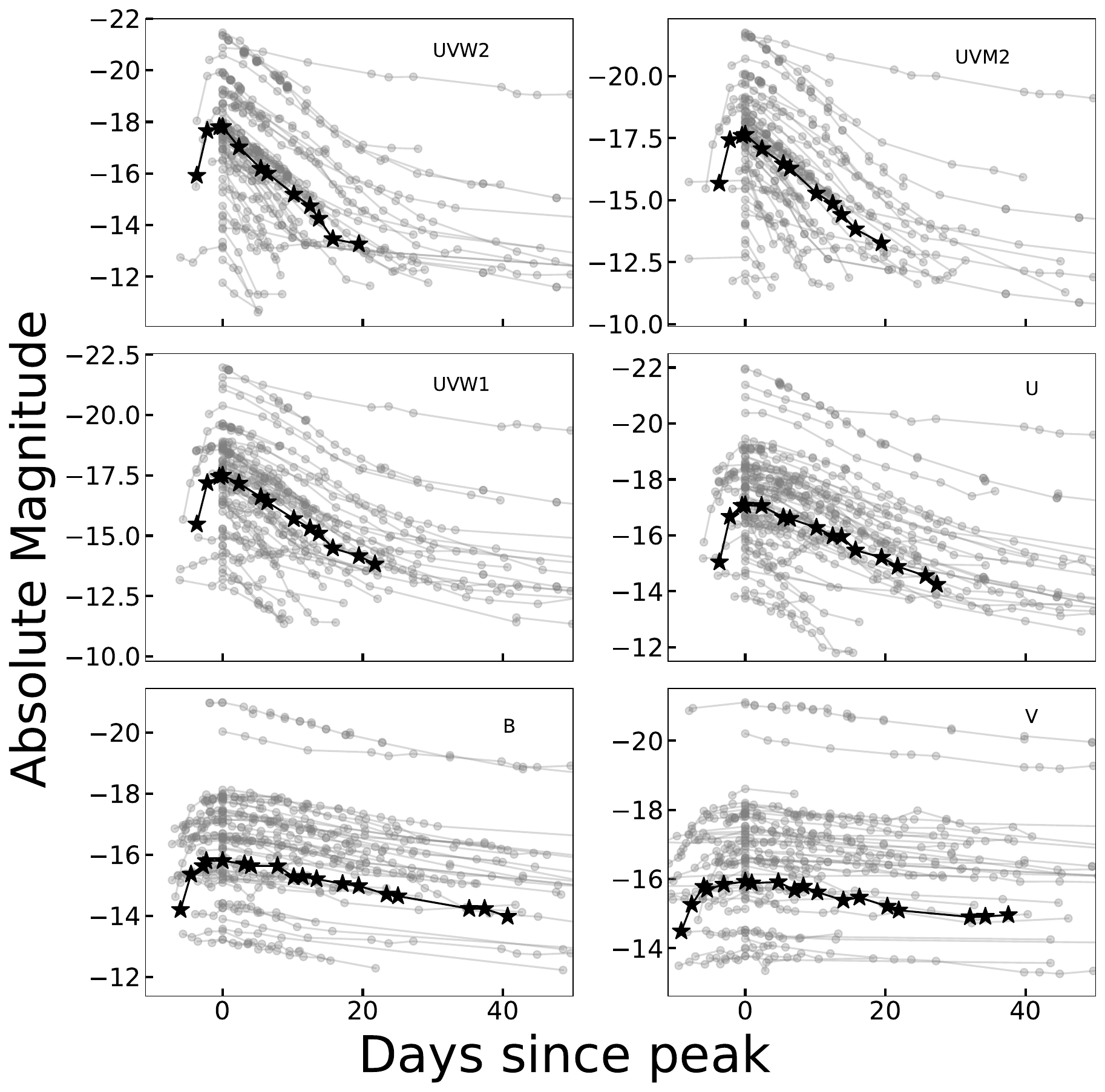}
    \caption{\textit{Top panel:} The absolute {\em V} band light curve of SN~2023zcu compared with the SNe of the comparison sample. The gray lines represent the light curves from \citet{Anderson_2014}. The vertical lines at 80 and 120 d represent the typical plateau duration for normal-plateau SNe. SN~2023zcu has a peak absolute magnitude of $-16.54\pm0.01$ in the {\em V} band, with a plateau length of $\sim$100 d, placing it in the category of intermediate luminosity normal-plateau SN. \textit{Bottom panel:} The \textit{Swift}/UVOT absolute magnitude light curves of SN~2023zcu (black stars) compared with a Type II SNe sample (grey points) taken from the \texttt{SOUSA}. All UVOT magnitudes are in  Vega, corrected only for Milky Way extinction.}
    \label{fig:absmag}
\end{figure}

SN~2023zcu has a normal-plateau duration and an intermediate luminosity. We compare its photometric characteristics with several other well-studied SNe that display a normal-plateau duration (e.g., SNe~1999em, 2006bp, 2009N, 2013fs, and 2023axu). In addition, to examine the behavior of the SN within the broader diversity of Type~IIP SNe, a SN with a short-plateau duration (SN~1995ad, which also exploded in the same host galaxy), a SN with a long-plateau duration (SN~2021yja), and a low-luminosity SN (SN~2005cs) are also selected to constitute the comparison sample. Some SNe (SNe~2006bp, 2021yja, and 2023axu) in the comparison sample are also selected because they exhibit similar early spectral features (flash ionization signatures) as observed in SN~2023zcu. The properties of the Type~IIP SNe in the comparison sample, along with SN~2023zcu, are summarized in Table~\ref{tab:comparison_objects}. The {\em V} band absolute peak magnitude of SN~2023zcu is similar to that of SN~2006bp. SN~2023zcu exhibits $M_{V50}$ of $-16.13\pm0.02$ and falls into the intermediate-luminosity SN category. The SN shows a photospheric phase lasting around 100 d, like SN~2023axu \citep[101.2$\pm$0.11;][]{Shrestha_2023axu}. The {\em V}  decline rate of the plateau is 0.84$\pm$0.01 mag (100 d)$^{-1}$, similar to that of SN~2009N \citep[0.84 mag (100 d)$^{-1}$;][]{takats_2009N, Valenti_2016}. The {\em V} band absolute magnitude light curve of SN~2023zcu is compared with other SNe in the sample (Figure~\ref{fig:absmag}, top Panel). The plateau duration for normal-plateau SNe \citep[80--120 d;][]{Nagy_2014, Anderson_2014, Arcavi_2012} is shown by two vertical lines. SN~2023zcu lies within this range, indicating a normal plateau length. 

After the plateau, the light curve declines at $8.18\pm0.09$ mag (100 d)$^{-1}$, followed by the nebular phase. \cite{Utrobin_2007} proposed that at the onset of the nebular phase, SNe are not fully powered by radioactive decay; instead, residual radiation energy flows from the hotter inner regions towards the cooler, transparent outer layers. This causes some luminosity excess, leading to a flattening of the initial nebular phase light curve known as the `plateau–tail', which typically lasts up to $\sim$26 days after the steep decline ends. During this phase, the decline rate in {\em V} band ($\gamma_V$) is slower than the $^{56}$Co decay rate \citep[0.98 mag (100 d)$^{-1}$,][]{Colgate_1969}. The plateau tail has been identified in several Type~IIP SNe, e.g., in SNe~1999em \citep[$\gamma_V$ = 0.93 mag (100 d)$^{-1}$,][]{Elmhamdi_2003, Utrobin_2007} and 2009N \citep[$\gamma_V$ = 0.45 mag (100 d)$^{-1}$,][]{takats_2009N}, where this phase lasts for $\sim$ 20 days and $\sim$ 40 days, respectively. It is even more evident in low-luminosity events, such as SNe~1999eu \citep[$\gamma_V$ = 0.07 mag (100 d)$^{-1}$,][]{Pastorello_2004} and 2005cs \citep[$\gamma_V$ = 0.46 mag (100 d)$^{-1}$,][]{Pastorello_2005cs_2009}, in which the plateau tail persists for about seven and six months after the onset of the nebular phase, respectively. For such a scenario, a P-Cygni profile is also observed in the spectral evolution over a period of almost one year. However, this type of flattening in the early nebular phase is not detected in SN~2021yja \citep[$\gamma_V$ = 1.39 mag (100 d)$^{-1}$,][]{Hosseinzadeh_2021yja_2022}. At later epochs, when the contribution from residual radiative energy vanishes, and the ejecta is powered entirely by radioactive decay, the luminosity decline becomes steeper. For SN~2023zcu, the plateau-tail decline rate (measured from the onset of the nebular phase over the first $\sim$ 25 days) is $\gamma_V$ = 0.81$\pm$0.26 mag (100 d)$^{-1}$, which is slower than the $^{56}$Co decay rate. This implies that at the onset of the nebular phase, the SN is not solely powered by radioactive energy. The radiative energy contributes additionally to the luminosity, resulting in a flattening of the light curve.

\begin{table*}
\centering
\renewcommand{\arraystretch}{1.5} 
    \caption{Properties of the Type IIP SNe in the comparison sample.}
    \label{tab:comparison_objects}
    \resizebox{\linewidth}{!}{%
    \begin{tabular}{ccccccccccc}
    \hline 
    Supernova & Parent & Distance & $A^{\rm tot}_V$ & $M_{V50}$ & $t_\mathrm{PT}$ & $E$ & $R$ & $M_\mathrm{ej}$ & $^{56}$Ni & Ref. \\
    & Galaxy & (Mpc) & (mag) & (mag) & (days) & ($10^{51}$~erg) & (R$_\odot$) & (M$_\odot$) & (M$_\odot$) &  \\

    \hline

    1995ad & NGC 2139 & 26.4 & 0.093 & $-16.6$ & $\sim$50 & 0.2 & 574.96 & 5.0 & $\sim$0.025 & 1\\

    1999em & NGC 1637 & 11.7(0.1) & 0.31 & $-$16.69$\pm$0.01 & 95 & $0.5-1$ & $120-150$ & $10-11$ & $0.042^{+0.027}_{-0.019}$ & 2, 3, 4 \\

    2005cs & M51 & 7.1(1.2) & 0.16 & $-$14.83$\pm$0.10 & $\sim$130 & 0.3 & -- & 8--13 & 0.003 & 5\\

    2006bp & NGC 3953 & 17.5$\pm$0.8 & 0.4 & - & - & $\sim$1 & - & $\sim$10 & - & 6\\ 

    2009N & NGC 4487 & 21.6(1.1) & 0.13 & $-15.35\pm0.01$ & 108.98$\pm$0.08 & $\sim$0.48 & $\sim$287 & $\sim$11.5 &0.020$\pm$0.004 & 7\\

    2013fs & NGC 7610 & 50.6$\pm$0.9 & 0.11 & $-16.78\pm0.01$ & 82.7$\pm$0.5 & 0.8–1.0 & - & 13.5–14.0 & 0.0545$\pm$0.0003 & 8 \\

    2021yja &  NGC 1325 &  21.8 & 0.32 & $-$17.5 & $\sim$125 & 1.53 & 631 & 15 & 0.175-0.2 & 9, 10\\

    2023axu & NGC 2283 & 13.68$\pm$2.05 & 1.23 & $-$16.85 & 101.2$\pm$0.3 & -- &  417$\pm$28 & 1.2$\pm$0.1 & 0.058$\pm$0.017 & 11\\ 

    \bfseries 2023zcu & \bfseries NGC 2139 & \bfseries 	27.8(2.0) & \bfseries 0.093 & \bfseries $-$16.13$\pm$0.02 & \bfseries 100.57$\pm$0.59 & $\mathbf{1.94^{+0.06}_{-0.06}}$ & $\mathbf{554.84^{+8.62}_{-8.62}}$ & 
    $\mathbf{10.01^{+0.34}_{-0.10}}$ & \bfseries  $\mathbf{0.023\pm0.004}$ & \bfseries This work \\  
    \hline
    \end{tabular}
    }
    References: (1) \cite{Inserra_1995ad_2013}, (2) \cite{Hamuy_2001}, (3) \cite{Leonard_2002}, (4) \cite{Elmhamdi_2003}, (5) \cite{Pastorello_2005cs_2009}, (6) \cite{Dessart_2006bp},(7) \cite{takats_2009N},  (8) \cite{Bullivant_2013fs}, (9) \cite{Kozyreva_2021yja_2022}, (10) \cite{Hosseinzadeh_2021yja_2022}.
    (11)\cite{Shrestha_2023axu}

\end{table*}

The \textit{Swift}/UVOT absolute magnitude light curves of SN~2023zcu are also compared with those of other Type IIP SNe observed with \textit{Swift} (Figure~\ref{fig:absmag}, bottom Panel). The sample light curves are downloaded from the \textit{Swift} Optical Ultraviolet Supernova Archive \citep[\texttt{SOUSA}\footnote{\url{https://archive.stsci.edu/prepds/sousa/}};][]{Brown_2014}, and are corrected only for Milky Way extinction. SN~2023zcu is not a UV-luminous SN, indicating that there is low CSM interaction during the early phase of the SN evolution.

\begin{figure}
    \centering
    \includegraphics[width=\linewidth]{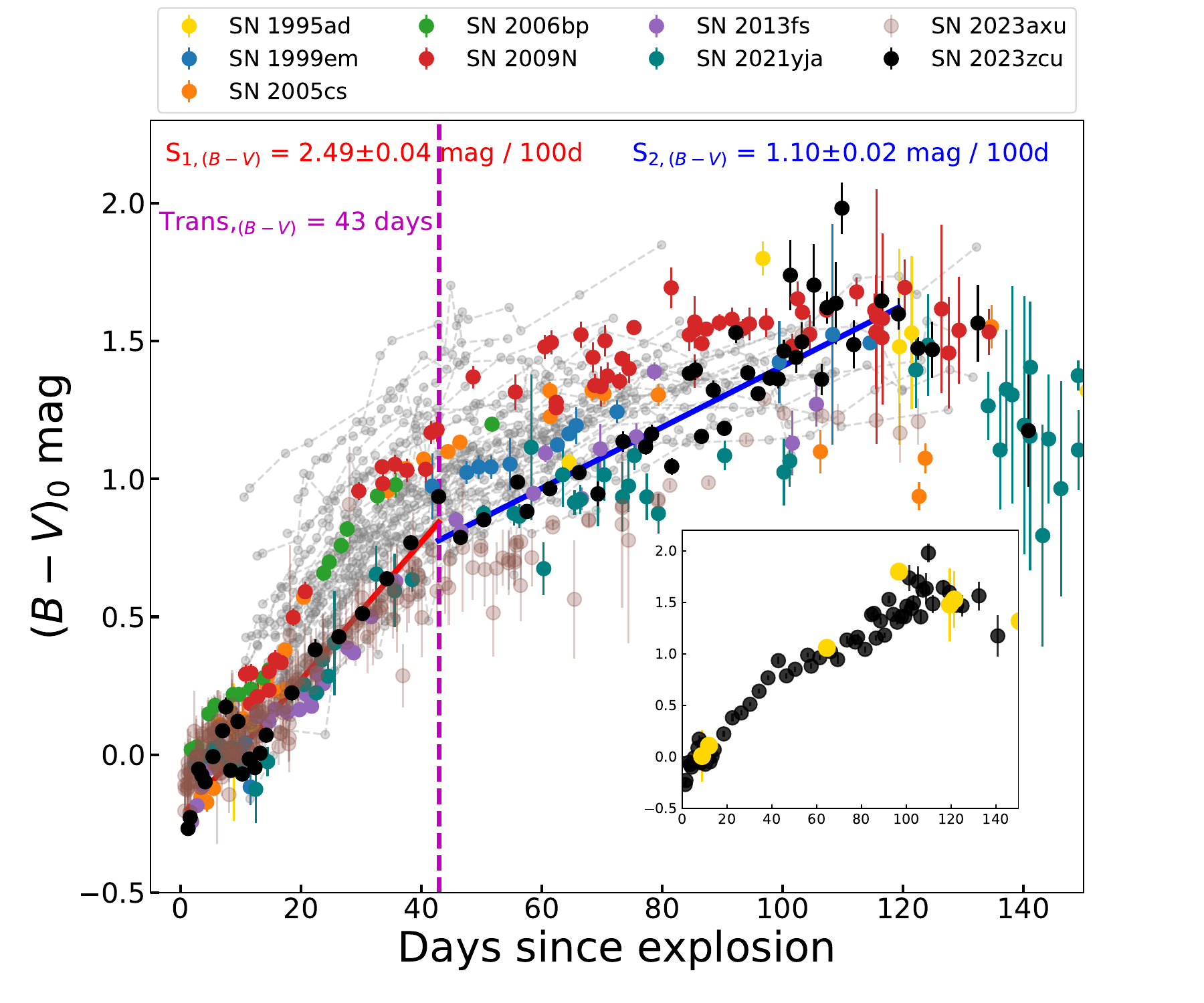}
    \caption{The $(B-V)_0$ color of SN~2023zcu is compared with the Type IIP SNe sample constructed for the analysis. The two slopes, S$_{1, (B-V)}$ and S$_{2, (B-V)}$, and the transition epoch between these two slopes, T$_{trans,(B-V)}$ (vertical line), are indicated in the figure. The colors of SNe~1995ad and 2023zcu are also shown in the inset. The gray points are the colors for the SNe sample taken from \cite{Jaeger_2018}.}
    \label{fig:b-v}
\end{figure}

\begin{figure*}
    \centering
    \includegraphics[width=1.0\linewidth]{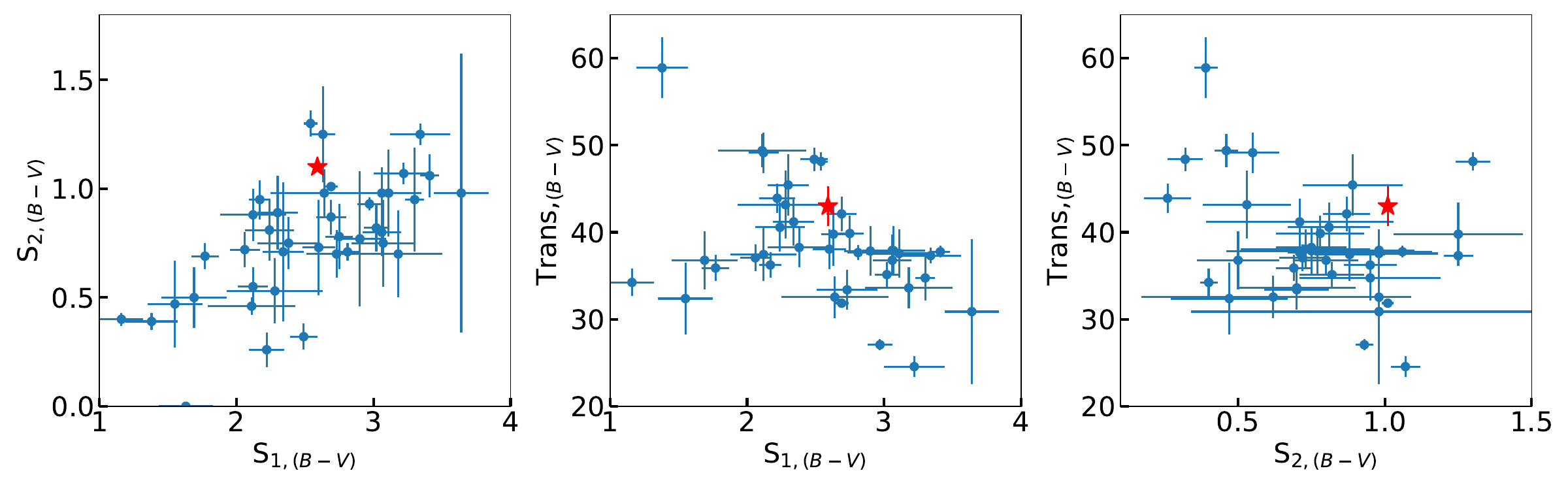}
    \caption{The correlations between the parameters, S$_{1, (B-V)}$, S$_{2, (B-V)}$ and T$_{trans,(B-V)}$, of the intrinsic $(B-V)_0$ color curve are presented. S$_{1, (B-V)}$ and S$_{2, (B-V)}$ are in units of mag (100 d)$^{-1}$ and T$_{trans,(B-V)}$ is in days. SN~2023zcu is highlighted with a red star. The other samples (marked in blue) are taken from \cite{Jaeger_2018}.} 
    \label{fig:b-v_compare}
\end{figure*}

The evolution and comparison of $(B-V)_0$ color is shown in Figure~\ref{fig:b-v}. The color evolution of SN~2023zcu is similar to that of SNe~1995ad, 2013fs, and 2021yja. As Type IIP SNe exhibit an approximately constant temperature during the plateau phase, they are expected to exhibit similar intrinsic color evolution, allowing extinction to be constrained from their observed colors \citep{Schmidt_1992, Olivers_2010}. In the inset plot of Figure~\ref{fig:b-v}, the colors of SNe~1995ad and 2023zcu are shown, since both these SNe exploded in the same galaxy. Their color curves are very similar, indicating a comparable extinction. The color curve of Type IIP SNe can be described with two slopes during the plateau phase. The first steeper slope, S$_{1,(B-V)}$, persists on average up to $\sim$40 d from the explosion, and the second relatively shallow slope, S$_{2,(B-V)}$, continues until the end of the plateau ($\sim$80 d) \citep{Jaeger_2018}. The transition period, denoted by T$_{trans,(B-V)}$, represents the time when the transition between S$_{1,(B-V)}$ and S$_{2,(B-V)}$ occurs. In SN~2023zcu, the $(B-V)_0$ color curve is first rising with a slope of S$_{1,(B-V)}$ = $2.49\pm0.04$ mag (100 d)$^{-1}$ until the transition at  T$_{trans,(B-V)}$ $\approx$ 43 d. After that, the color gradually evolves with a slope of S$_{2,(B-V)}$ = $1.10\pm0.02$ mag (100 d)$^{-1}$.  In Figure~\ref{fig:b-v_compare}, the color parameters are compared in the correlation plots with the sample taken from \cite{Jaeger_2018}. The first plot reveals a positive correlation between S$_{1,(B-V)}$ and S$_{2,(B-V)}$, indicating that SNe, which cool more rapidly at early epochs, also tend to cool more quickly during the plateau phase. SN~2023zcu cools at an average rate in the early epoch, whereas in the latter phase, it cools faster than most SNe. 

\subsection{$^{56}$Ni Mass Estimation}
\label{sec:Nimass}

In the nebular phase, the photosphere reaches the base of the SN, and the ejecta become optically thin. At this stage, the SN is entirely powered by the radioactive decay of $^{56}$Ni$\rightarrow$$^{56}$Co$\rightarrow$$^{56}$Fe. The $\gamma$ rays are down-scattered by the SN ejecta, giving rise to emission in the optical bands. Several methods have been developed to constrain the amount of synthesized $^{56}$Ni during SN explosions; in this analysis, the following approaches are employed.

\begin{enumerate}
    \item One way to estimate the $^{56}$Ni mass is by assuming that the SN has the same spectral energy distribution (SED) as the well-studied SN~1987A. For SN~2023zcu, the decay rate in the nebular phase is $s_3$ = 0.99$\pm$0.01 mag (100 d)$^{-1}$, which is close to the $^{56}$Co decay slope (0.98 mag (100 d)$^{-1}$), suggesting complete trapping of $\gamma$ rays. The $^{56}$Ni mass of SN~1987A ($0.075\pm0.005$ M$_\odot$) was accurately determined in \cite{Arnett_1996}. Hence, by scaling the bolometric luminosity of the SN during the nebular phase to that of SN~1987A, the $^{56}$Ni mass can be determined \citep{Spiro_2014}. Using this method, the $^{56}$Ni mass is calculated at six nebular epochs (132.43, 140.82, 154.12, 249.83, 301.40, and 403.43 d). The weighted average of these values, including the distance uncertainty in quadrature, yields 0.026$\pm$0.014 M$_\odot$.
    
    \item Another similar approach is to estimate the $^{56}$Ni mass through the radioactive tail luminosity using bolometric corrections ({$BC$}) given by \cite{Hamuy_2003}. In this method, the {\em V} band tail photometry is used, implemented with $BC$ \citep{hamuy2001PhD} to get the bolometric luminosity. The $^{56}$Ni mass is obtained at the same six nebular epochs used in the previous approach. To accurately estimate the error in $^{56}$Ni mass, we employ a Bayesian framework. We compare the $V_t$ values calculated from the method provided by \cite{Hamuy_2003} to the observed ones, using the posterior probability

    \begin{equation}\label{prob}
    \begin{split}
    P(M_{Ni} \mid V_t) 
    &= P\big(V_t \mid M_{\rm Ni}, t_0, z, D, BC\big) \\
    &\quad \times P\big(M_{\rm Ni}, t_0, z, D, BC\big)\,
    \end{split}
    \end{equation}
\noindent
     The first term on the right-hand side is the log-likelihood, and the second term is the prior. For $t_0$, $z$, $D$ (in cm), and $BC$, we assume Gaussian priors characterized by their value and uncertainties, and for $^{56}$Ni mass ($M_{\rm Ni}$), we assume a flat prior between 0 and 0.1. Since $t_0$, $z$, $D$, and $BC$ have well-constrained priors, their posterior just reflects the priors, but we are able to constrain the posterior of $M_{\rm Ni}$ well. We run 10,000 iterations of MCMC simulations and find $M_{\rm Ni} = 0.023 \pm 0.003$ M$_\odot$. 
\end{enumerate}

\noindent
The weighted average of the estimates from the above two methods is adopted as the final mass of $^{56}$Ni = $0.023\pm0.004$ M$_\odot$.

\section{Spectroscopic Analysis}
\label{sec:spectroscopic_analysis}

\subsection{Early-Time Spectroscopy}

\begin{figure}
    \centering
\includegraphics[width=\linewidth]{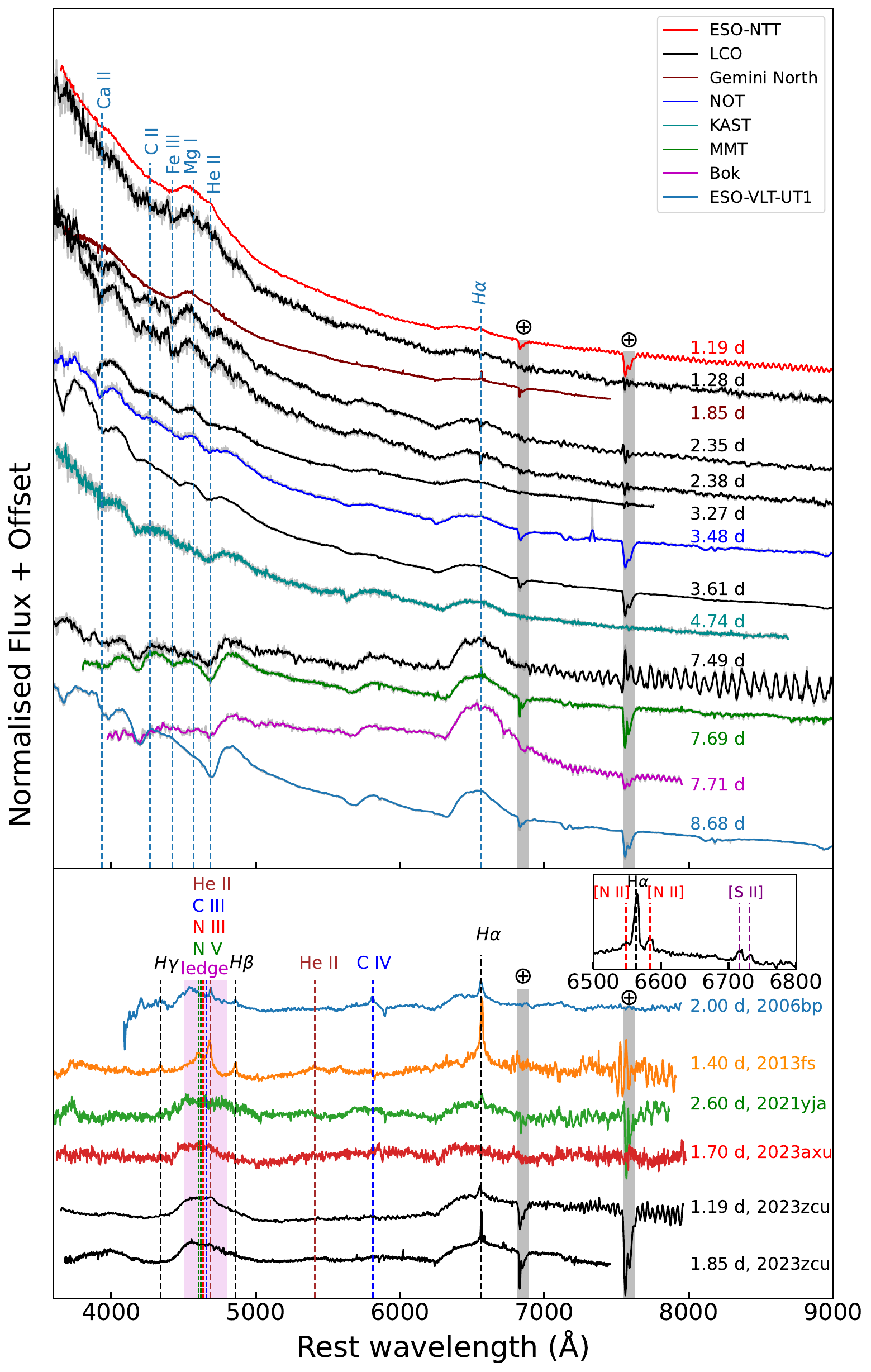}
    \caption{\textit{Top panel:} The early-phase spectral evolution of SN~2023zcu from 1.19 to 8.68 d is shown. All spectra are redshift-corrected and normalized with respect to the median values. \textit{Bottom panel:} The comparison of the 1.19 and 1.85 d spectra of SN~2023zcu with four SNe of the comparison sample, showing early time CSM interaction signatures. The host galaxy contamination is present in the 1.85 d spectrum, as narrow emission lines originating from the galaxy ([\ion{S}{2}] 6717, 6731 \AA\ and [\ion{N}{2}] 6548, 6584 \AA) are visible (see inset). The ledge-shaped feature around 4500–-4800 \AA, highlighted in pink color, is possibly a blend of multiple weak high-ionization lines, such as \ion{He}{2}, \ion{C}{3}, \ion{N}{3}, and \ion{N}{5}.}
    \label{fig:early_spec}
\end{figure}

The redshift-corrected and normalized spectra during the early phase, spanning from 1.19 to 8.68 d (in the observer frame of reference), are shown in Figure~\ref{fig:early_spec} (top panel). All spectra are smoothed. A broad and weak H$\alpha$ emission profile with little to no absorption is visible over the blue continuum until 4.74 d. A faint \ion{He}{2} feature is observed in the first spectrum and disappears thereafter. Other emission lines, such as \ion{Mg}{2}, \ion{Fe}{3}, \ion{C}{2}, and \ion{Ca}{2} are also marked for reference. High-ionization emission lines, typically observed in some Type II SNe during the early phase, due to the ionization of CSM caused by the shock breakout of the SN \citep{Smith_2015, Khazov_2016, Bruch_2021, Bruch_2023}, are absent in the early spectra of SN~2023zcu. However, a weak emission line on top of the broad H$\alpha$ P-Cygni emission component becomes visible in the 1.19 and 1.85 d spectra. This absence in the 1.28 d spectrum could be attributed to its low resolution, which can wash out any narrow emission component arising from the CSM with velocities of the order of 100 km s$^{-1}$. In the 1.85 d spectrum, the presence of narrow emission lines from the host galaxy ([\ion{S}{2}] 6717, 6731 \AA, [\ion{N}{2}] 6548, 6584 \AA) indicates host-galaxy contamination. 

Figure~\ref{fig:early_spec} (bottom panel) compares the 1.19 and 1.85 d spectra of SN~2023zcu with other SNe exhibiting CSM interaction. A narrow emission line, superimposed on the broad H$\alpha$ P-Cygni profile, is similarly observed in SNe~2006bp and 2021yja. While SN~2023zcu lacks other high-ionization emission lines such as \ion{He}{2}, \ion{C}{4}, \ion{N}{5}, and \ion{C}{3}, it displays a distinctive asymmetric `ledge-shaped' feature around 4500–-4800 \AA, as seen in SNe~2021yja \citep{Hosseinzadeh_2021yja_2022} and 2023axu \citep{Shrestha_2023axu}. In SN~2021yja, this ledge feature was explained as a result of early-stage low CSM interaction, which is insufficient to create high-ionization lines but enough to affect the broad-band light curves, mainly the luminous UV one \citep{Hosseinzadeh_2021yja_2022}. \cite{Bullivant_2013fs} described a similar feature in SN~2013fs as a broad and blueshifted component of \ion{He}{2} emission arising from the SN ejecta under the CSM. A similar feature is also noticed in SN~2006bp. The narrow \ion{He}{2} component in SN~2023zcu is not prominent and resembles that observed in SNe~2021yja and 2023axu. However, SN~2023zcu is not highly UV-luminous (see Figure~\ref{fig:absmag}, bottom panel), unlike SN~2021yja. On the other hand, this ledge feature can also be observed in non-UV-luminous events, such as SN~2023axu. It could arise from a possible blend of several weak high-ionization lines, including \ion{He}{2}, \ion{C}{3}, \ion{N}{3}, and \ion{N}{5} \citep{Soumagnac_2020, Bruch_2021, Hosseinzadeh_2021yja_2022}. Hence, the presence of a narrow H$\alpha$ line and ledge feature supports the idea of low-level CSM interaction with SN ejecta.

\begin{figure}
    \centering
    \includegraphics[width=\linewidth]{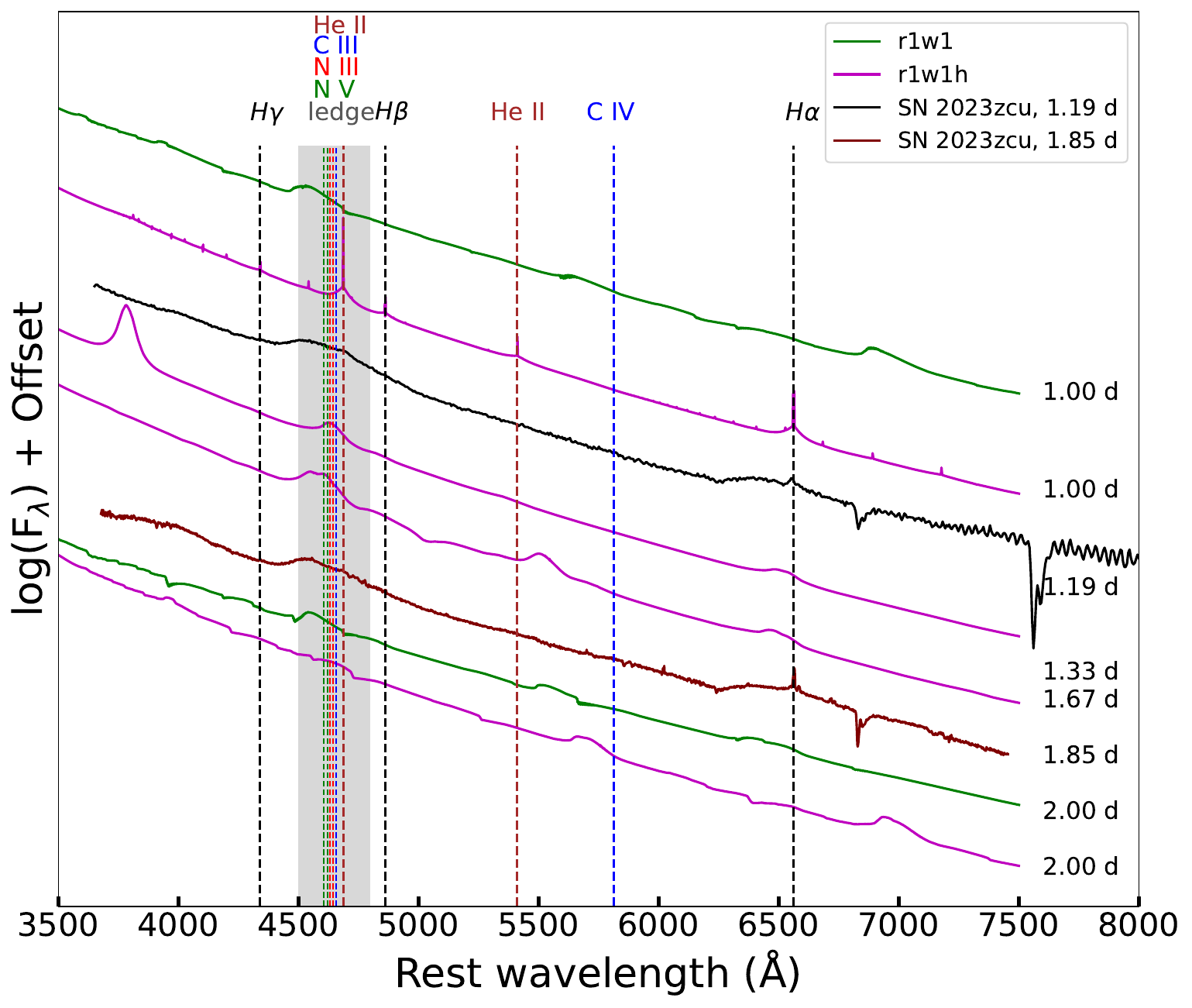}
    \includegraphics[width=\linewidth]{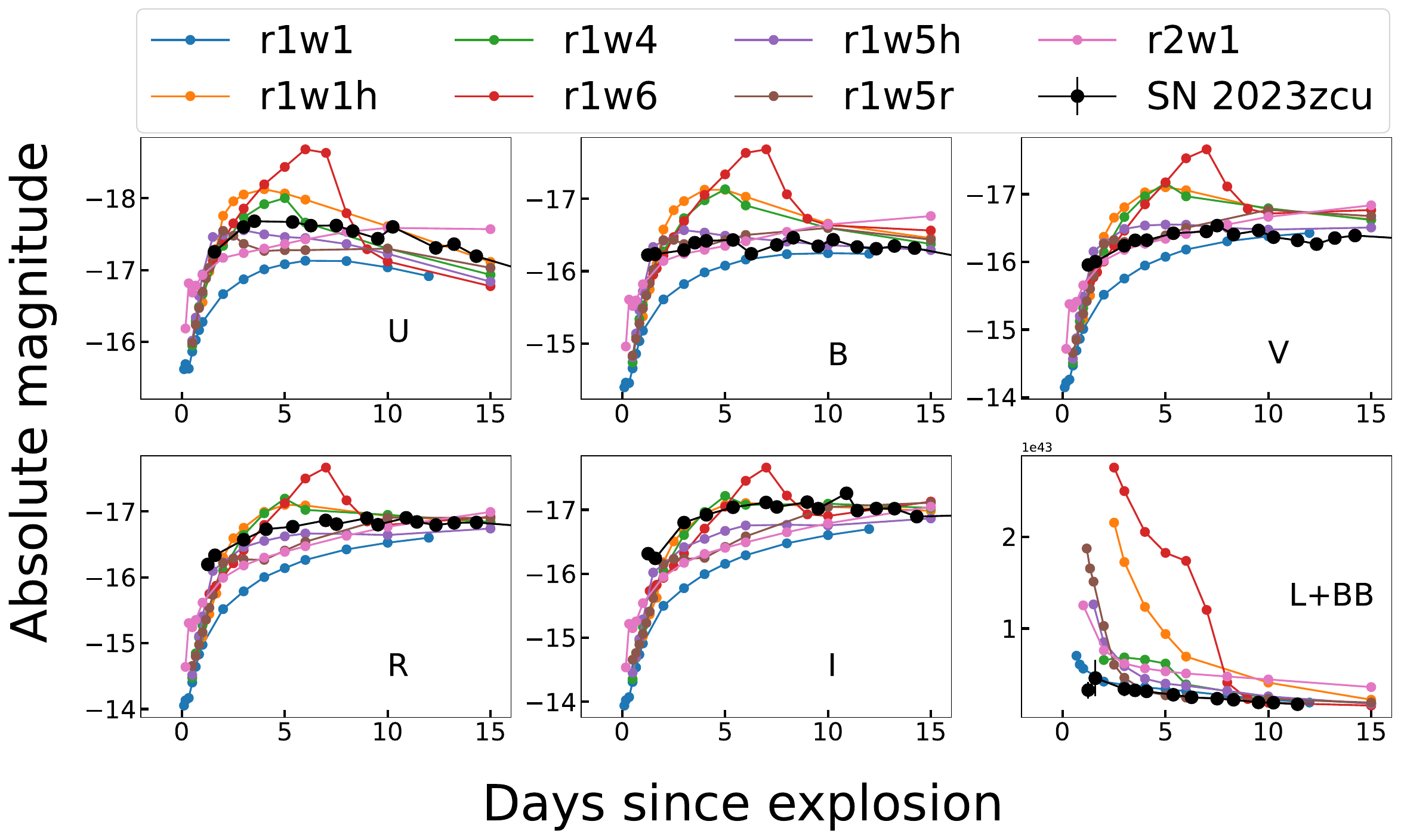}
    \caption{\textit{Top panel:} The 1.19 and 1.85 d spectra of SN~2023zcu are compared with the \texttt{r1w1} and \texttt{r1w1h} models from \cite{Dessart_2017}, including the identification of high-ionization lines. \textit{Bottom panel:} The multiband absolute magnitude and bolometric light curves of SN~2023zcu are compared with the models of \cite{Dessart_2017}.}
    \label{fig:dessert_model}
\end{figure}

\cite{Dessart_2017} produced synthetic spectra of CCSNe using CMFGEN, a one-dimensional non-local thermodynamic equilibrium (NLTE) radiative-transfer model \citep{Dessert_CMFGEN_2012}. The model (\texttt{r1w1}) with a compact ($R$ = 501 R$_\odot$) progenitor with the lowest CSM mass-loss rate (10$^{-6}$ M$_\odot$ yr$^{-1}$) produces the ledge kind of feature; however, narrow emission lines are not visible or may disappear a few hours after the explosion. In this model, they added an extended atmosphere onto the RSG (\texttt{r1w1h} model; scale height $H_\rho = 0.3 \times 501$ R$_\odot$), which produces a similar kind of spectroscopic signature, brighter magnitude in the UV, and in the early phase a narrow H$\alpha$ emission line. The terminal velocity of the wind is considered to be 50 km s$^{-1}$ in both of the models. In Figure~\ref{fig:dessert_model} (top panel), the first two early spectra are compared with the \texttt{r1w1} and \texttt{r1w1h} models. A ledge feature around 4500--4800 \AA\ can be seen in both the models. The H$\alpha$ narrow emission, along with additional emission lines such as H$\beta$, \ion{He}{2}, and \ion{C}{3}, are only produced by the earliest spectrum (at 1.00 d) of model \texttt{r1w1h}. In Figure~\ref{fig:dessert_model} (bottom panel), {\em UBVRI} magnitudes and the bolometric luminosity of SN~2023zcu, generated with \texttt{SuperBol} \citep{Nicholl_2018} using \textit{Swift}/UVOT and {\em UVBri} bands, are compared with the \cite{Dessart_2017} models. During the early phase, the \texttt{r1w1h} model is brighter than the observed magnitude of SN~2023zcu in the {\em UVB} bands, whereas the {\em RI} magnitudes are closely matched. It explains that SN~2023zcu was not UV-luminous at the early phase owing to low CSM interaction. The observed magnitudes are brighter than the \texttt{r1w1} model in all five bands, whereas the bolometric luminosity is similar to the \texttt{r1w1} model. These comparisons suggest that the progenitor could be an RSG star with a less extended atmosphere than assumed in model \texttt{r1w1h}, embedded within low-mass CSM, which is intermediate to the values assumed in both models.

\subsection{Plateau and Nebular Phase Spectroscopy}

\begin{figure}
    \centering
    \includegraphics[width=\linewidth, height=2.1\linewidth]{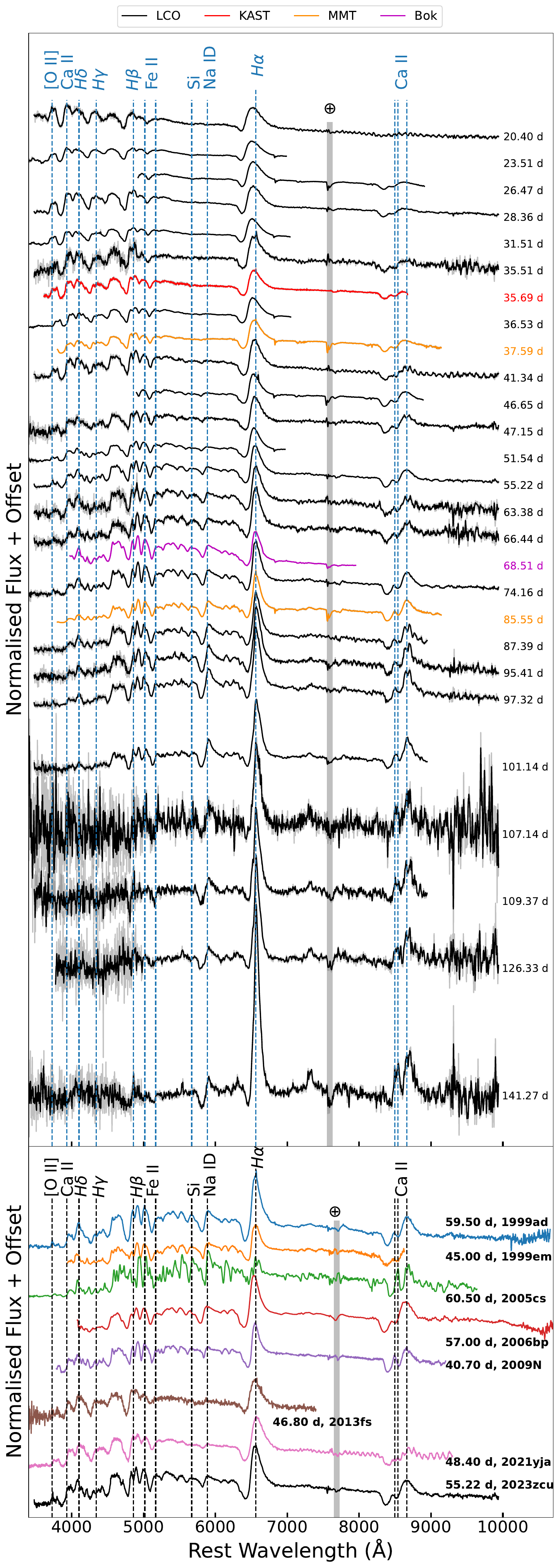}
    \caption{\textit{Top panel:} Spectroscopic evolution of SN~2023zcu from 20.40 to 141.27 d. All spectra are redshift-corrected and normalized with respect to the median values. Spectral features are marked with dotted lines. \textit{Bottom panel:} The mid-plateau-phase spectrum of SN~2023zcu at 55.22 d is compared with SNe of the comparison sample at similar epochs. The H$\alpha$ profile closely matches that of well-defined Type IIP SNe, which also exhibit well-developed metal lines.}
    \label{fig:plateau_spectra}
\end{figure}

Figure~\ref{fig:plateau_spectra} (top panel) presents the spectral evolution of SN~2023zcu from the early plateau (20.40 d) to the early nebular phase (141.27 d) in the observer frame of reference. All spectra are redshift-corrected, normalized with respect to median values, and smoothed. The early plateau spectra contain several emission lines at bluer wavelengths, whereas the line profiles at redder wavelengths become prominent in the later spectra. The P-Cygni profile of the H$\alpha$ line is prominent in each spectrum, suggesting the existence of an extended H envelope in the progenitor prior to the explosion. The emission counterpart is stronger than the absorption of the H$\alpha$ profile. A weak H$\beta$ 4861.3 \AA\ line is possibly blended with \ion{He}{2} 4686 \AA\ at early phases, and then becomes prominent from 35.51 d. The \ion{Na}{1~D} 5890, 5896 \AA\ doublet is visible from 41.34 d and becomes prominent in later spectra, indicating the cooling of the SN. The metal lines like \ion{Fe}{2} 5018, 5169 \AA, \ion{Ca}{2} near-infrared (NIR) triplet (8498, 8542, 8662 \AA) become prominent as the SN evolves. The spectra become noisy during the onset of the nebular phase (107.14 d). As the ejecta are converting to an optically thin medium, H$\alpha$ emission becomes narrower and more prominent than the absorption component. In Figure~\ref{fig:plateau_spectra} (bottom panel), the 55.22 d spectrum of SN~2023zcu is compared with the sample SNe constructed for comparison at similar epochs. The spectral signatures closely resemble those of the sample SNe, except for SN~2005cs, whose H$\alpha$ profile is not as strong as the others, which is expected since this is a low-luminosity Type IIP SN. Weak [\ion{O}{2}] 3727 \AA\ emission is detected in both SNe~1995ad and 2023zcu, which likely originates from the host galaxy. The metal lines, such as \ion{Fe}{2} and the \ion{Ca}{2} NIR triplet, are still developing in SN~2021yja, whereas they are prominent in SN~2023zcu, similar to other SNe in the sample.

\begin{figure}
    \centering
    \includegraphics[width=\columnwidth]{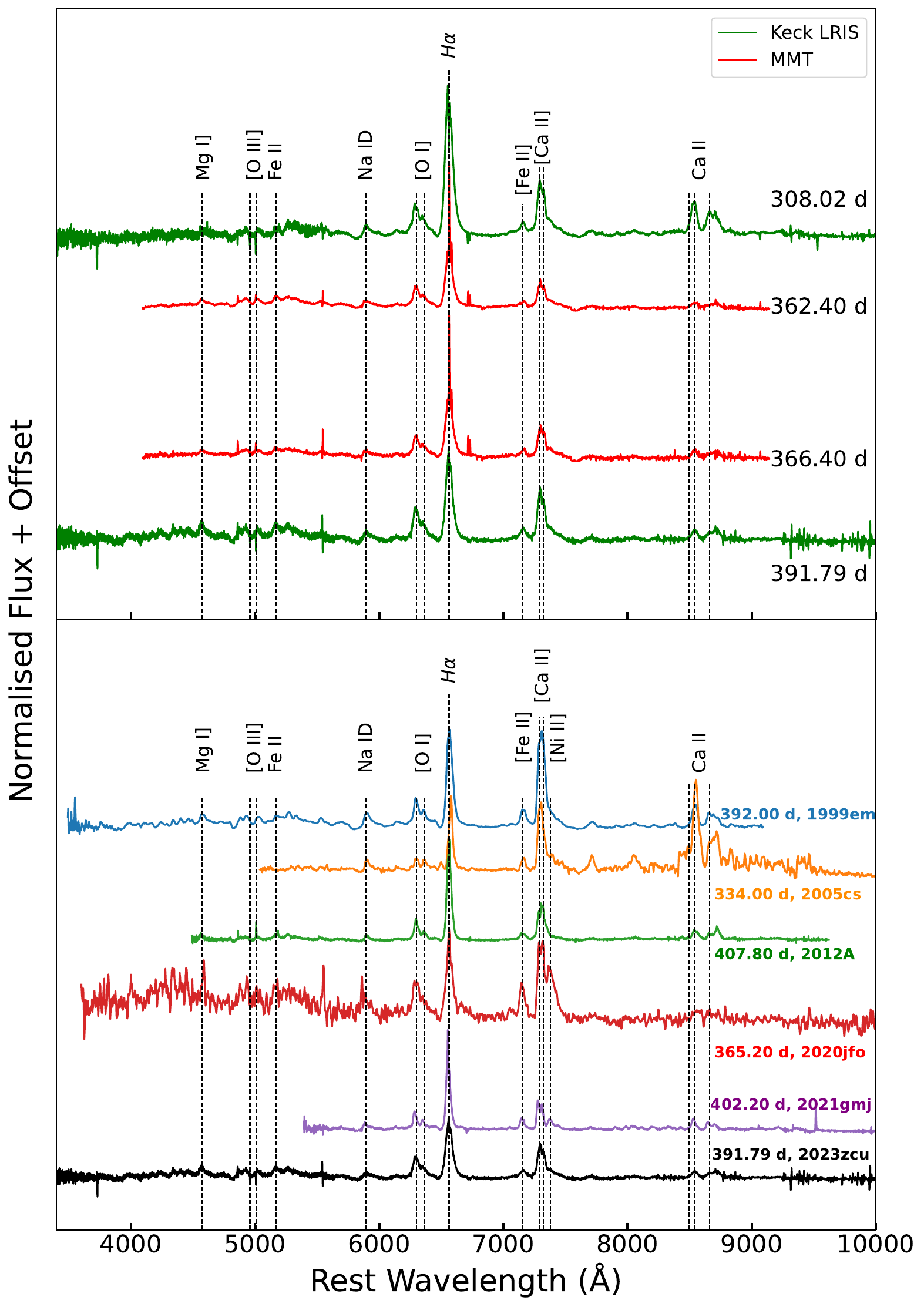}
    \caption{\textit{Top panel:} The spectral evolution of SN~2023zcu in the nebular phase is shown. All spectra are redshift corrected and normalized with respect to the median values. Spectral features are marked with dotted lines. \textit{Bottom panel:} The nebular spectrum at 391.79 d is compared with five sample SNe. The spectral features closely resemble those of the Type IIP SNe~1999em, 2012A, and 2021gmj. Prominent [\ion{Ni}{2}] 7378 \AA\ next to the [\ion{Ca}{2}] line is detected in SNe~2020jfo and 2021gmj, but is not visible in SN~2023zcu.}
    \label{fig:neb_spectra}
\end{figure}

During the late nebular phase, the SN ejecta are almost transparent to optical photons, providing the opportunity to probe deeper into the ejecta. The redshift-corrected and normalized four late-time nebular spectra are obtained between 308.02 and 391.79 d (in the observer frame of reference), presented in Figure~\ref{fig:neb_spectra} (top panel). A strong H$\alpha$ emission line is visible because of the recombination of H in the ejecta, which is ionized by the radioactive-decay energy. In the low-density environment, several forbidden lines, such as [\ion{Ca}{2}] 7291, 7324 \AA, [\ion{O}{1}] 6300, 6364 \AA, and weak \ion{Mg}{1}] 4571 \AA, arise owing to transitions between metastable energy levels. \ion{Fe}{2}, \ion{Na}{1~D}, and the \ion{Ca}{2} NIR triplet lines are also visible, revealing the metal-rich ejecta. In Figure~\ref{fig:neb_spectra} (bottom panel), the 391.79 d spectrum is compared to SNe~1999em and 2005cs from the comparison sample and three additional SNe~2012A \citep{tomosella_2012A}, 2020jfo \citep{Ailawadhi_2023}, and 2021gmj \citep{Meza-Retamal_2021gmj_2024, Murai_2021gmj_2024}. The H$\alpha$ emission line and the [\ion{O}{1}] doublet are similar to SNe~1999em, 2012A, 2020jfo, and 2021gmj, unlike for the low-luminosity SN~2005cs, where the [\ion{O}{1}] lines are much weaker than the other nebular lines. SN~2023zcu exhibits weaker \ion{Ca}{2} than SN~2005cs, but similar to the other SNe in the sample. There is a prominent [\ion{Ni}{2}] 7378 \AA\  redward of the [\ion{Ca}{2}] line detected in SNe~2020jfo and 2021gmj, which is not visible in SN~2023zcu.

\subsection{Spectral Modeling with \textsc{TARDIS}}
\label{sec:tardis}

To estimate the physical parameters of SN~2023zcu in the early photospheric phase, radiative-transfer modeling of the spectra taken from 8.68 to 35.51 d after the explosion using a modified version of the radiative-transfer code \textsc{TARDIS} \citep{Kerzendorf_2014}, repurposed for the early photospheric phase of Type II SNe \citep{Vogl_2019}, is performed. As described by \cite{Vogl_2019}, the code assumes the ejecta to be spherical and homologously expanding, described by a power-law density profile (in the form of $\rho = \rho_0(r/r_0)^{-n}$, where $\rho_0$ denotes the density at a characteristic radius $r_0$, and $n$ representing the power-law slope) and a uniform chemical composition. These assumptions are well-founded for the modeling of Type II SNe (\citealt{DessartHillier_2006, Dessart_2006bp, Vogl_2019}). Furthermore, \textsc{TARDIS} treats excitation and ionization under NLTE conditions, allowing for accurate modeling of the Balmer lines. On the other hand, it does not account for time-dependent effects and also neglects the NLTE treatment of iron-group elements, both of which become significant with metal line blanketing a few weeks after the explosion. Consequently, \textsc{TARDIS} Type II SN modeling is most applicable in the first month of the SN evolution. 

The spectral emulator presented by \cite{Vogl_2020} is utilized to fit the spectral time series of SN~2023zcu, using the model grid detailed by \cite{Csoernyei_2023b} for the training of the emulator. The spectral fitting was performed on a maximum-likelihood basis, following the strategy presented by \cite{Vogl_2020}, \cite{Vasylyev_2023}, and \cite{Csoernyei_2023a}. As detailed in Section~\ref{sec:Dis_explosion}, given the lack of substantial \ion{Na}{1~D} absorption-line detection, and nominal color evolution, we assumed the \cite{Schlafly_2011} dust map foreground reddening estimate of $E(B-V) = 0.03$ mag for the spectral modeling. This is consistent with the reddening estimates for SNe~1995ad \citep{McNaught_1995ad} and 2022qhy \citep{Smith_2022qhy}. We offer the caveat of reddening due to dust local to the SN site; however, an extra reddening component within an uncertainty budget of 0.05 mag is generally undetectable by SN spectral modeling \citep{DessartHillier_2006}, and larger components are unlikely given the observed colors of SN~2023zcu. To correct the spectra for the assumed extinction, the \cite{Cardelli_1989} extinction law with a total-to-selective extinction ratio $R_V = 3.1$ is employed. For the fitting, the O$_2$ and H$_2$O band telluric regions have been masked. Any further corrections based on photometry are not employed, and the modeling is performed directly on the extracted spectra. To this end, only the LCO dataset, along with the VLT spectra, is modeled to ensure good quality of the flux calibration.

\begin{figure}
    \centering
    \includegraphics[width=1.0\linewidth]{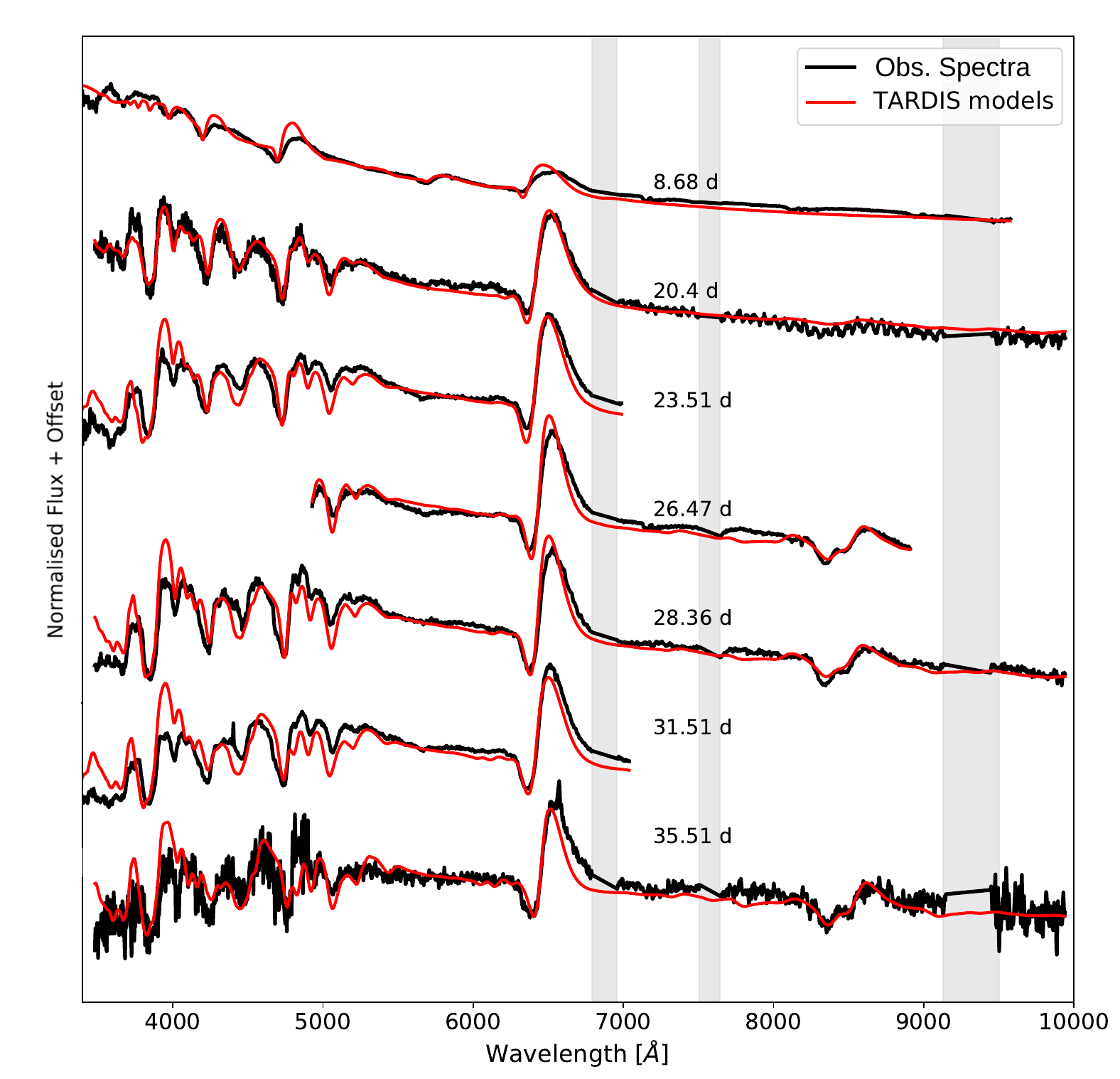}
    \caption{Best-fit \textsc{TARDIS} models (red) compared to the observed spectra (black). The gray shaded columns mark the telluric regions, which were masked and excluded from the fitting. As shown, the fits adequately reproduce the various SN features.}
    \label{fig:tardis}
\end{figure}

The best-fit models estimated for the preset reddening are displayed in Figure~\ref{fig:tardis}, while the resulting physical parameters are listed in Table~\ref{tab:tardis}. The good quality of the fits indicates that SN~2023zcu is within the realm of normal Type IIP SNe in terms of spectral evolution, with only a minimal amount of potential CSM interaction, in line with the discussion presented in previous sections. In general, the profile of the Balmer series is well reproduced, except for the first epoch, where the issue is similar to that encountered by \cite{Dessart_2006bp}, who observed that a very steep density profile must be assumed to properly reproduce the line profiles at such early phases. 

\begin{table}
    \centering
    \caption{Best-fit parameters obtained through \textsc{TARDIS} modeling.}
    \label{tab:tardis}
    \begin{tabular}{cccc}
        \hline
        Epoch (days) & $n$ & $v_{\textrm{ph}}$ [km/s] & $T_{\textrm{ph}}$ [K] \\
        \hline
         8.68 & 17.04 & 10076.65 & 9246.73\\
         20.41 &	9.06 & 7584.53 & 6906.29\\
         23.52 &	8.08 &	7395.34 & 5998.17\\
         26.47 &	8.23 &	6122.78 & 5998.24\\
         28.36 & 8.18 &	6464.29 & 5998.13\\
         31.52 &	8.02 &	7067.22 & 5998.16\\
         35.52 &	7.71 &	5415.86 & 5998.31\\

        \hline
    \end{tabular}
    
\end{table}

\subsection{Tailored-EPM Distance Estimation}
\label{Tailored_EPM}

Type II SNe enable the estimation of distances independently of the cosmic distance ladder through variants of the EPM \citep {Kirshner1974, Dessart_2005, Vogl_2024}. This method is based on the assumption that in the early stage of the SN evolution, with fully ionized ejecta, the opacity is governed by electron scattering at the photosphere, which can be considered radiating as a diluted blackbody. The observed flux and color temperature yield the angular radius of the photosphere when corrected for a wavelength-dependent dilution factor \citep{Schmidt_1992}. The photospheric expansion velocity ($V_{\rm ph}$), measured from the spectra, provides the photospheric radius through $R = v_{\rm ph} \times (t-t_0)$, where $t_0$ is the explosion epoch. By comparing the angular and photospheric radii, the distance to the SN can be derived \citep{Krishner_1994, hamuy2001PhD, Hamuy_2002}. As demonstrated by \cite{Csoernyei_2023a} and \cite{Vogl_2024}, Type II SNe are now capable tools for inferring high-precision distances with only a limited need for computationally intensive processes. Following the spectral modeling outlined in Section~\ref{sec:tardis}, which is the necessary input for the tailored-EPM method, we estimated the distance of SN~2023zcu and hence NGC~2139 following the recipe outlined by \cite{Csoernyei_2023a}.

The method is centered around the determination of the photospheric angular diameter of the SN ($\Theta = R_{\textrm{ph}}/ D$, where $R_{\textrm{ph}}$ and $D$ denote the radius of the photosphere and the distance, respectively). This quantity is inferred for each of the spectral epochs based on the \textsc{TARDIS} spectral models, by minimizing the difference between measured and model apparent magnitudes ($m^{\textrm{obs}}$ and $m$, respectively) estimated from the fits at the given epoch, using $\Theta$ as argument

\begin{equation}
\label{eq:EPM_mod}
    \Theta^{*} = \arg \min_{\hspace*{-15pt}\Theta} \sum_{\textrm{S}}\left(m_{\textrm{S}} - m_{\textrm{S}}^{\textrm{obs}}\right)^2
\end{equation}

\noindent for all $S$ photometric bands available. To estimate the model apparent magnitudes for each of these bands, one has to employ the distance-modulus formula and replace the distance with the angular diameter,

\begin{equation}
\label{eq:EPM}
\begin{split}
    &m_S - M_S = - 5 + 5 \log (D) + A_S\, ,\\
    &m_S = M_S - 5 + 5 \log \frac{R_{\textrm{ph}}(\Sigma^{*})}{\Theta(\Sigma^{*})} + A_S\, ,\\
    &m_S = M_S^{\textrm{ph}}(\Sigma^{*}) - 5 \log [\Theta(\Sigma^{*})] + A_S\, , \\
\end{split}
\end{equation}

\noindent as detailed by \cite{Csoernyei_2023b}. Here, $\Sigma^{*}$ corresponds to the set of best-fitting physical parameters, $M_S^{\textrm{ph}}$ is the absolute magnitude at the position of the photosphere, $D$ is in pc, and $A_S$ denotes the dust extinction. With this setup, by estimating $M_S^{\textrm{ph}}(\Sigma^{*})$ through the models and by interpolating the $m_{\textrm{S}}^{\textrm{obs}}$ observed magnitudes for the spectral epochs utilizing the light-curve fitting from \cite{Csoernyei_2023a}, a $\Theta$ value for each spectrum is estimated. For the resulting $\Theta / v_{\textrm{ph}}$ values, Gaussian uncertainties of 10\% following \cite{DessartHillier_2006} are assumed.

\begin{figure}
    \centering
    \includegraphics[width=1\linewidth]{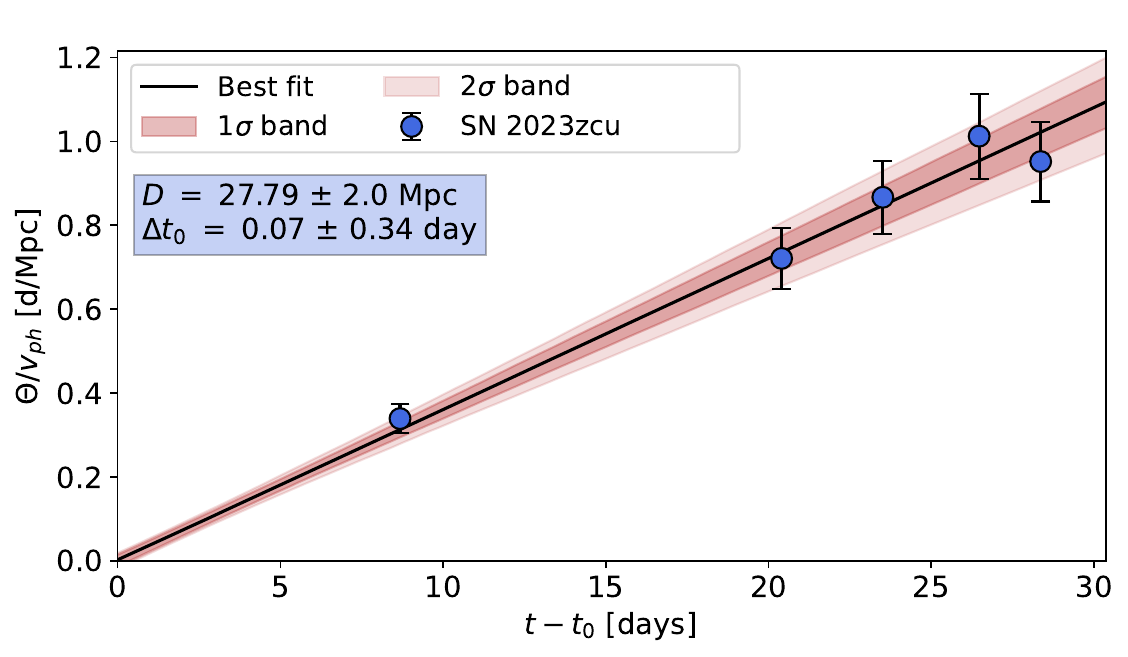}
    \caption{EPM regression applied to the data of SN~2023zcu. The shaded regions denote the 1$\sigma$ and 2$\sigma$ uncertainty bands. The abscissa shows the time elapsed since the explosion. The resulting distance and $t_0$ correction implied by the fitting are displayed in the blue box.}
    \label{fig:2023zcu_EPM}
\end{figure}

Finally, we combine these values with the assumption of homologous expansion $R_{\textrm{ph}} = v_{\textrm{ph}}\times (t-t_0)$ (which is well motivated by observations and models for normal Type II SNe; see, e.g., \citealt{Woosley_1988, Dessart_2005}). This allowed us to determine the distance to the SN and its time of explosion through a Bayesian linear fit to the ratios of the angular diameters and the photospheric velocities ($\Theta / v_{\textrm{ph}}$) against time, $t$. For the fitting, a flat prior is used in both $D$ and $t_0$ (time of explosion), the latter of which was set as the time between the last nondetection and the discovery of SN~2023zcu (Table~\ref{tab:SN 2023zcu and Host information}). The resulting EPM regression is displayed in Figure~\ref{fig:2023zcu_EPM}. For the EPM distance estimation, only the spectra younger than 30 d are used, and the first five fits are displayed in Figure~\ref{fig:tardis}, to limit our exposure to systematic uncertainties arising from neglecting time-dependent effects in the underlying \textsc{TARDIS} simulations (see discussions by \citealt{Vogl_2020} and \citealt{Csoernyei_2023b}).

The final distance estimate obtained using the LCO $BgVri$ bands is $D = 27.8 \pm 2.0$ Mpc. The EPM regression did not allow for further specification of $t_0$, owing to the lack of well-calibrated spectra at earlier epochs. The obtained distance is within the range of the Tully-Fischer distances available for this galaxy, which are 17--37 Mpc as listed in NED (e.g., \citealt{Bottinelli_1985, Mathewson_1992, Theureau_2007}), with a mean distance of $26.8\pm1.2$ Mpc as noted in Table~\ref{tab:SN 2023zcu and Host information}. Furthermore, this estimate is also in line with the standardizable candle method (SCM) value of $26.40 \pm 2.57$ Mpc obtained by \cite{Pejcha_2015} for the sibling SN~1995ad. The good consistency highlights the potential and precision of tailored-EPM for distance estimation, whose precision can be further improved to the few-percent level when additional spectra are available in the early photospheric phase.

\begin{figure}
    \centering
    \includegraphics[width=1\linewidth]{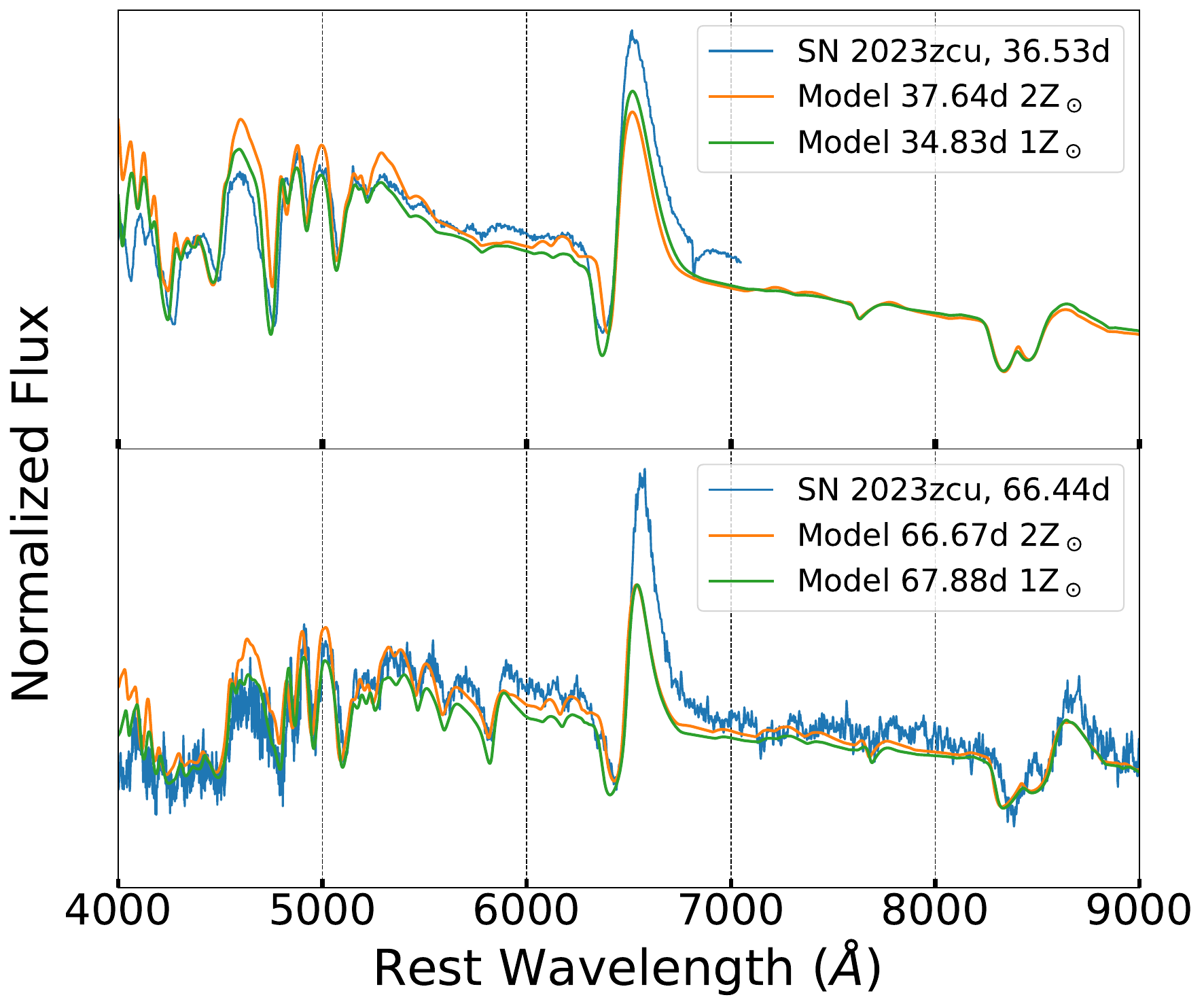}\\
    \includegraphics[width=1\linewidth]{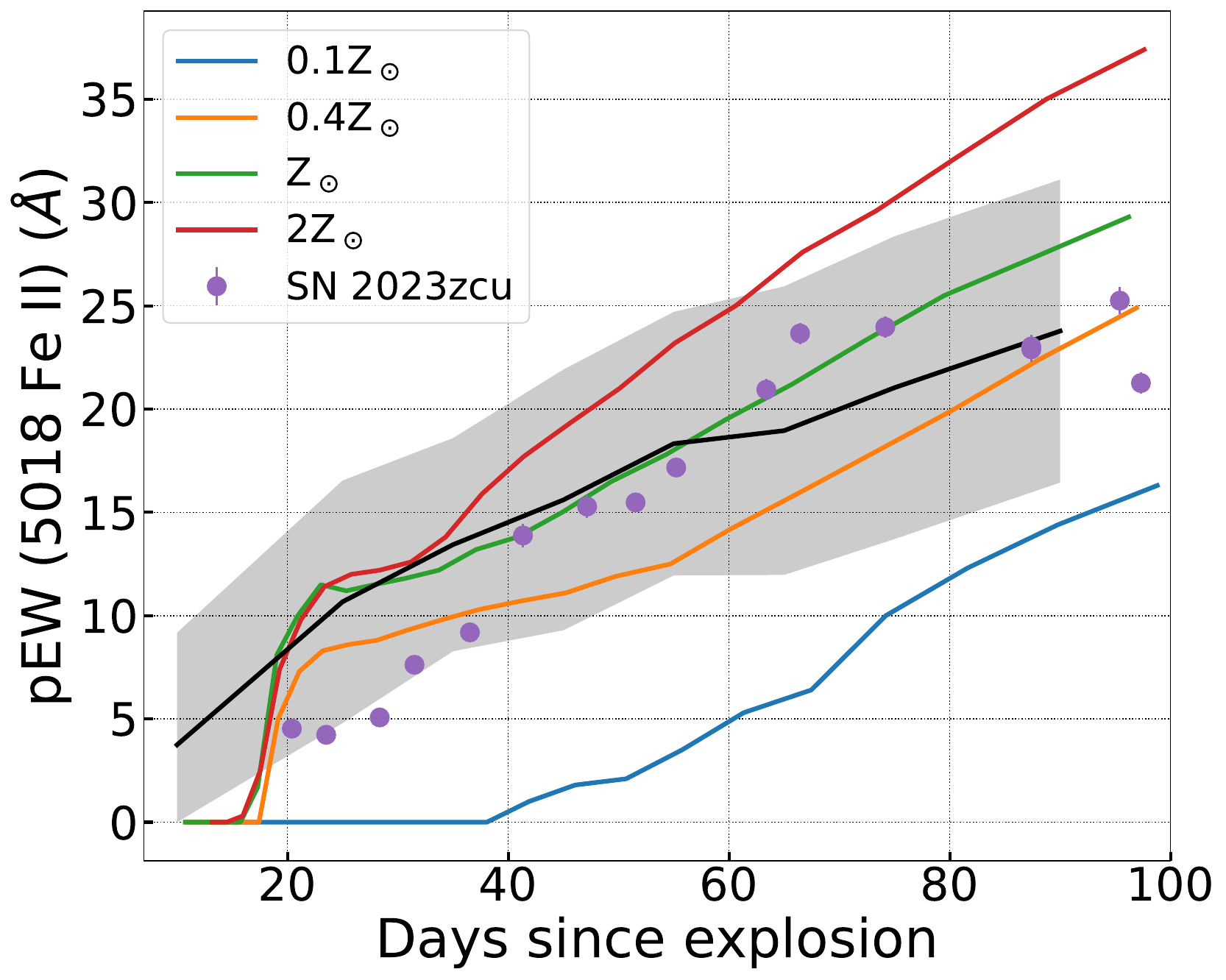}
    \caption{\textit{Top panel:} Two plateau-phase spectra (36.53 and 68.51 d) are overplotted with model spectra from \citet{Dessart_2013} at similar epochs, generated with 1\,Z$_\odot$ and 2\,Z$_\odot$ models. Both models are equally good comparisons to the observed spectra.
    \textit{Bottom panel:} The time evolution of the pEW of the \ion{Fe}{2} 5018 \AA\ line of SN~2023zcu is shown. The black solid line represents the time evolution of the mean pEW from 119 Type II SNe from \citet{Anderson_2016}, with uncertainties presented by the gray-shaded region. The pEW calculated from the model spectra for different metallicities presented by \citet{Dessart_2013} are plotted in distinct colors. SN~2023zcu has near solar metallicity.}
    \label{fig:metallicity}
\end{figure}

\subsection{Estimation of Metallicity}
\label{metallicity_vel}

The estimation of metallicity is crucial for understanding the local host-galaxy environment and, therefore, for constraining the progenitor properties \citep{Dessart_2014}. \cite{Inserra_1995ad_2013} derived the metallicity of the local environment for SN~1995ad by measuring  O3N2 and N2 indices \citep{Pettini_2004} of an \ion{H}{2} region close to the SN. The resulting metallicity, 12 + log(O/H) $= 8.60 \pm 0.05$ dex, is very close to solar metallicity \citep[8.65 dex;][]{Asplund_2009}. To estimate the metallicity of SN~2023zcu, two spectra taken during the plateau phase (36.53 and 66.44 d) are compared with the model spectra \citep{Dessart_2013} generated for different metallicities (0.1, 0.4, 1.0, 2.0 Z$_\odot$). In Figure~\ref{fig:metallicity} (top panel), the spectra from models 1\,Z$_\odot$ and 2\,Z$_\odot$ are overplotted with SN~2023zcu spectra at two similar epochs. The \ion{Fe}{2} 5018, 5169 \AA\ feature is well matched in both the models and observed spectra at both epochs.

\cite{Anderson_2016} studied the oxygen abundance of the \ion{H}{2} regions of 119 SN host galaxies through emission-line fluxes, and obtained a positive correlation with the pseudo-equivalent width (pEW) of the \ion{Fe}{2} 5018 \AA\ line at 50 d post-explosion spectra of the corresponding SNe. They found Type II SNe to be useful as metallicity indicators; however, progenitor metallicity is not a key factor driving the diversity in SN light curves and spectra, except for the pEW of the \ion{Fe}{2} 5018 \AA\ line. In Figure~\ref{fig:metallicity} (bottom panel), the time evolution of the mean pEW of 119 SNe (black line) along with the uncertainty (gray shaded region) from \cite{Anderson_2016} is shown. The measured pEW of the \ion{Fe}{2} 5018 \AA\ for different spectral epochs of SN~2023zcu is plotted, along with the pEW of model spectra from \cite{Dessart_2013} generated at different metallicities, represented in different colors. To calculate the pEW, firstly, the noisy spectra were smoothed. Then four regions were defined, continuum regions on the left and right side, and regions with mark the start and end of the absorption dip. 100 Monte Carlo iterations are performed, where each iteration consists of randomly picking one point from each region, defining the continuum and the absorption dip, and then calculating the pEW. We report the median value as the pEW, along with the 1$\sigma$ errors. The pEW of the \ion{Fe}{2} 5018 \AA\ line of SN~2023zcu is similar to the 1\,Z$_\odot$ model, indicating that SN~2023zcu exhibits solar metallicity.

\begin{figure}
    \centering
    \includegraphics[scale=0.32]{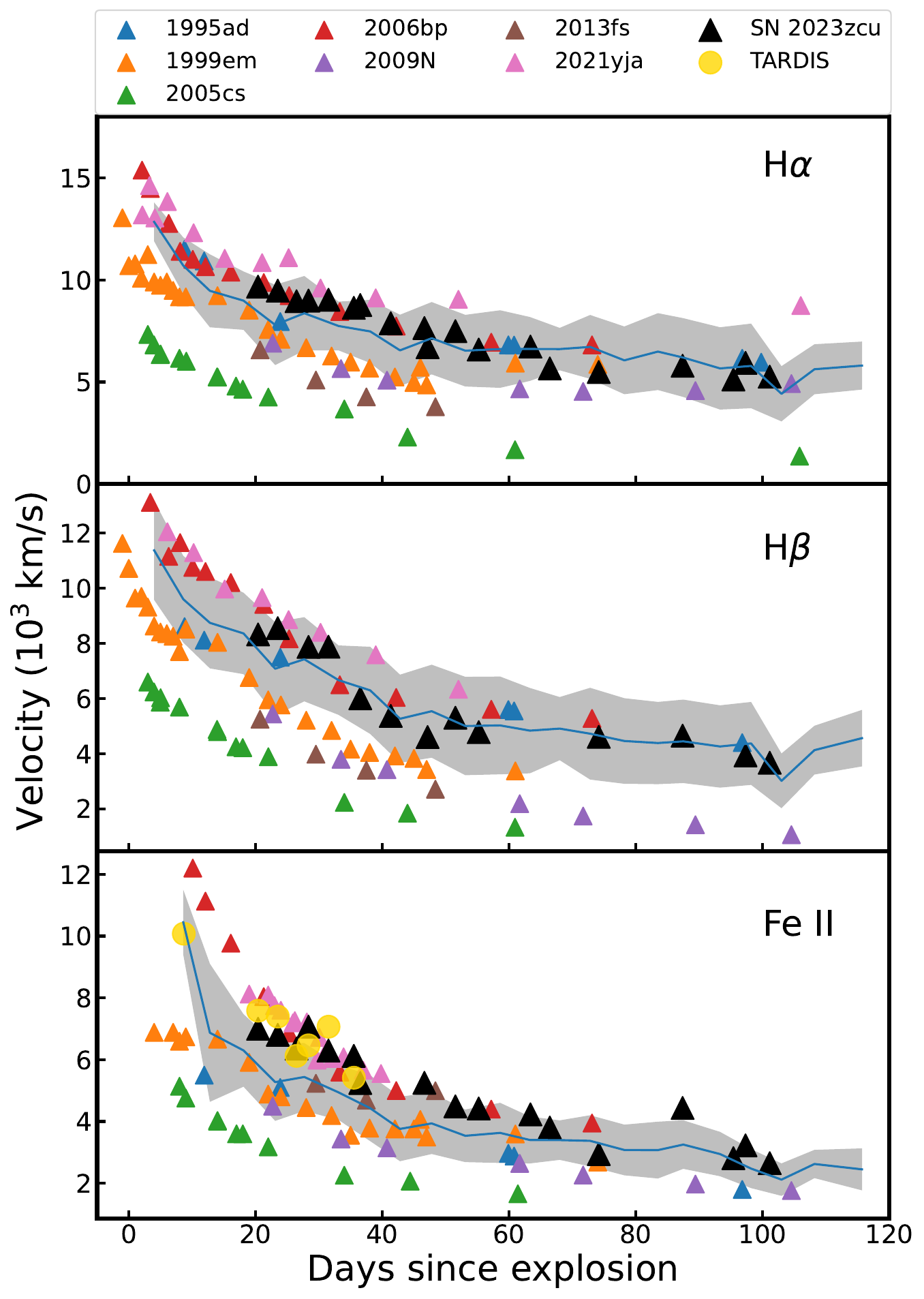}\\
    \includegraphics[scale=0.34]{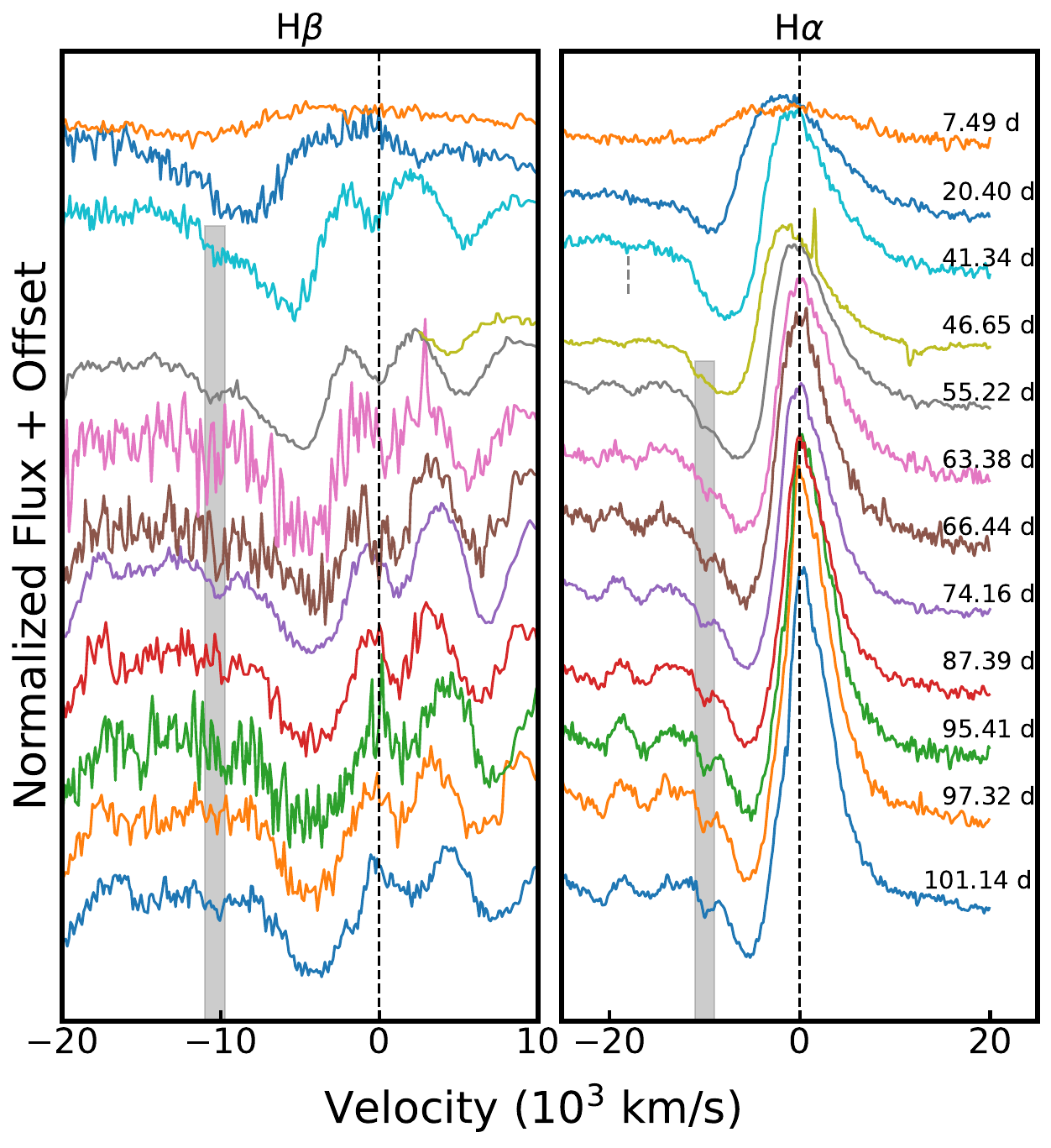}
    
    \caption{\textit{Top panel:} Velocity evolution of H$\alpha$, H$\beta$, and \ion{Fe}{2} 5169 \AA\ lines in SN~2023zcu are compared with the SNe in the comparison sample. The blue solid line and the gray shaded region represent the mean velocities and the corresponding standard deviations of 122 Type II SNe presented in \cite{Gutierrez_2017}. Yellow circles in the \ion{Fe}{2} velocity plot represent the velocities obtained from \textsc{TARDIS} modeling. \textit{Bottom panel:} The evolution of the `Cachito' feature \citep{Gutierrez_2017} blueward of the H$\alpha$ profile during the plateau phase. This feature becomes prominent in the latter phases, with its counterpart also seen blueward of H$\beta$. A small absorption feature blueward of H$\alpha$ is seen at 41.34 d. The gray shaded region represents velocities ranging from 9,000 to 11,000 km s$^{-1}$.}
    \label{fig:vel_evolution}
\end{figure}

\subsection{Velocity Evolution}
The velocities of three prominent lines (H$\alpha$, H$\beta$, and \ion{Fe}{2} 5169 \AA) are calculated from their blueshifted absorption minima. In Figure~\ref{fig:vel_evolution} (top panel), these velocities are compared with the SNe from the comparison sample and with the mean velocity of 122 SNe, along with their 1$\sigma$ uncertainty presented in \cite{Gutierrez_2017}. The H$\alpha$ line velocity is higher than the H$\beta$ and \ion{Fe}{2} velocities, indicating that H$\alpha$ originates in the outer hydrogen-rich layers of the ejecta, whereas the H$\beta$ and \ion{Fe}{2} lines are formed in relatively deeper layers of the ejecta. The \ion{Fe}{2} velocity can be used as a proxy for the photospheric velocity, and it is estimated to be around 7,000 km s$^{-1}$ at 20 d and declines to $\sim$2,500 km s$^{-1}$ at the beginning of the nebular phase. In the early plateau phase ($t>35$ d), the line velocities are higher than the mean velocity, similar to SNe~2006bp and 2021yja. After that, the velocities decrease and are comparable to the mean velocity. The H$\alpha$ and H$\beta$ velocity evolution of SN~1995ad closely follows that of SN~2023zcu; however, its \ion{Fe}{2} 5169 \AA\ velocity is significantly lower. SN~2023zcu exhibits a higher velocity than the normal Type IIP SNe~1999em, 2009N, 2013fs, and the low-luminosity SN~2005cs. In the \ion{Fe}{2} velocity plot, the yellow circles represent the velocity obtained from the \textsc{TARDIS} modeling over a span of one month, and are well matched with the velocities obtained from the absorption minima.   

An extra absorption feature on the blue side of H$\alpha$ and H$\beta$ spectral lines is visible during the plateau phase (gray shaded regions in Figure~\ref{fig:vel_evolution}, bottom panel). This feature is referred to as the `Cachito' feature \citep{Gutierrez_2017}. \cite{Chugai_2007} proposed that this feature might have originated from the interaction between SN ejecta with the RSG wind or CSM, or from the ionization of the ejecta by the X-ray produced by the reverse shock; however, \cite{Pastorello_2006} identified it as the \ion{Si}{2} 6355 \AA\ line in SN~2005cs. For SN~2023zcu, the Cachito feature starts to appear blueward of H$\alpha$ from 46.75 d, while a shallow absorption dip is visible at 42.34 d. The velocity of this feature ranges from 9,000 to 11,000 km s$^{-1}$, which is similar to the early-time H$\alpha$ velocity (see Figure~\ref{fig:vel_evolution}, top panel). The presence of the corresponding counterpart blueward of H$\beta$ (which is shallow due to low optical depth at the line-forming region) is observed within the same velocity range \citep{Gutierrez_2017}, and its existence at later phases (50--100 d; \citealt{Chugai_2007}) would suggest an association with the high-velocity (HV) component of the H$\alpha$ profile. Therefore, in SN~2023zcu, this Cachito feature is most likely produced by interaction between the ejecta and the RSG wind or CSM.

\section{Progenitor Properties}
\label{sec:progenitor}

\subsection{Constraining Progenitor Mass from Nebular Spectroscopy}

\begin{figure}
    \centering
    \includegraphics[width=\columnwidth]{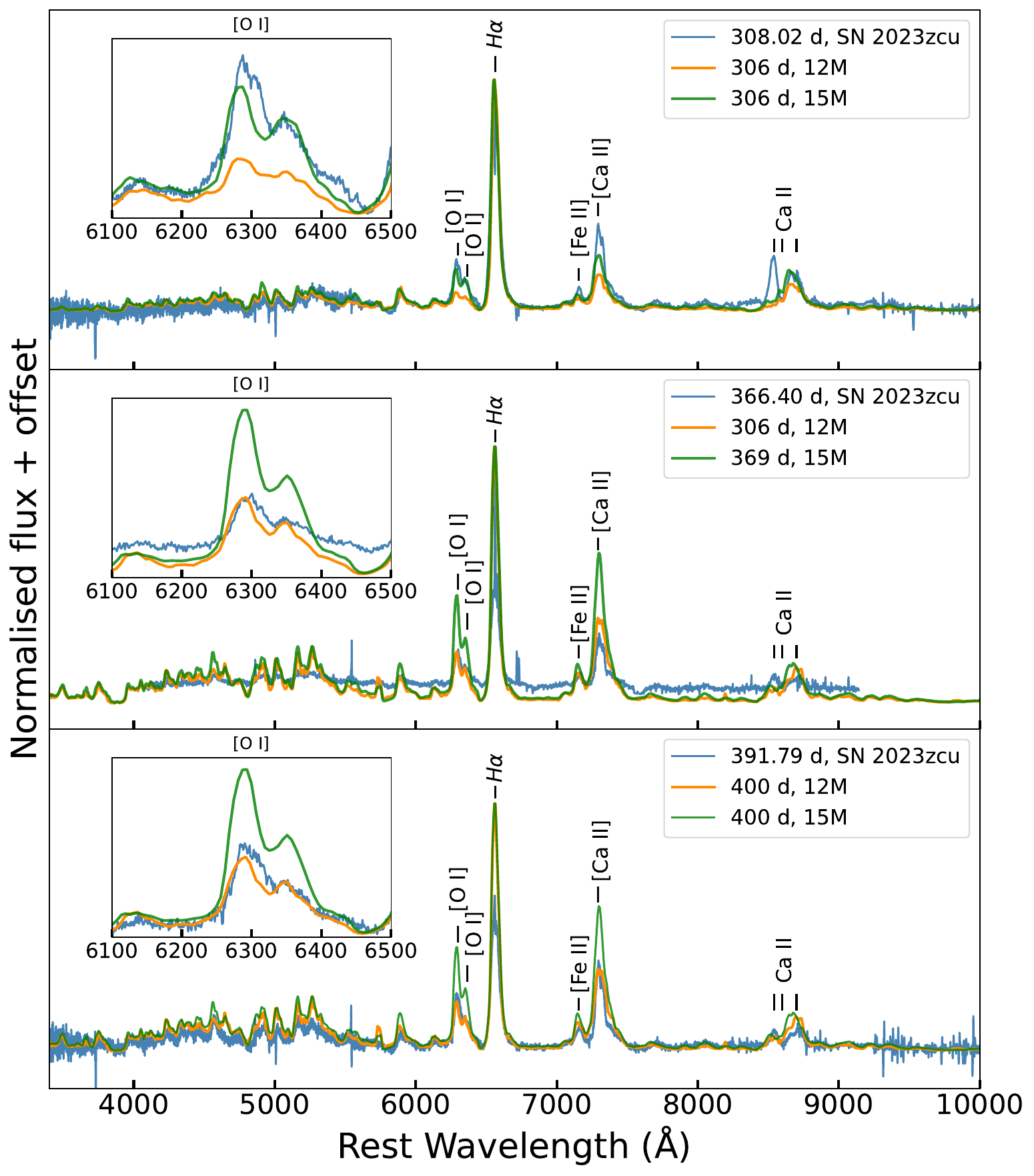}
    \caption{The nebular spectra of SN~2023zcu at 308.02 (upper panel), 366.40 (middle panel), and 391.79 d (lower panel) are compared with model spectra from \cite{Jerkstrand_2014}, computed for 12 and 15 M$_\odot$ progenitors at similar epochs. The inset highlights the [\ion{O}{1}] 6300, 6364 \AA\ doublet, a good indicator of progenitor mass, of both the observed and model spectra, suggesting a progenitor mass between 12 and 15 M$_\odot$.
    }
    \label{fig:JK_model}
\end{figure}

The strength of the nebular lines can give information about the progenitor mass, as the nucleosynthesis depends strongly on the main-sequence mass of the star \citep{Woosley_1995}. \cite{Jerkstrand_2012, Jerkstrand_2014} found in their studies that the strength of the [\ion{O}{1}] 6300, 6364 \AA\ doublet in the nebular spectra is proportional to the zero-age main-sequence (ZAMS) mass of the progenitor. For Type IIP SNe, \cite{Jerkstrand_2014} developed models for different progenitor masses (12, 15, 19, and 25 M$_\odot$) at multiple nebular epochs. They used the \texttt{KEPLER} code \citep{Woosley_2007} to evolve and explode the stars using the model described by \cite{Jerkstrand_2011} and generated synthetic spectra. Three nebular spectra of SN~2023zcu, obtained at 308.02, 366.40, and 391.79 d, are compared with the model spectra, computed for 12 and 15 M$_\odot$ progenitors at 306 and 400 d. The model spectra are scaled to the observed spectra to match the integrated flux using Equation~\ref{eu:neb_scalling} in \cite{Bostroem_2019},

\begin{equation} \label{eu:neb_scalling}
    \frac{F_{\rm obs}}{F_{\rm mod}} = \frac{d^2_{\rm mod}}{d^2_{\rm obs}} \times \frac{M_{Ni_{\rm obs}}}{M_{Ni_{\rm mod}}} \times \exp\left(\frac{t_{\rm mod} - t_{\rm obs}}{111.4}\right)
\end{equation}

\noindent   
Here, $F_{\rm obs}$ and $F_{\rm mod}$ are the total flux of the observed and model spectra. $d_{\rm obs}$ is the distance of the SN in Mpc and $d_{\rm mod}$ = 5.5 Mpc is used to compute the model. $M_{Ni_{\rm obs}}$ is the estimated $^{56}$Ni mass of the SN and $M_{Ni_{\rm mod}}$ is taken as 0.062 M$_\odot$, whereas $t_{\rm obs}$ and $t_{\rm mod}$ are the epochs of observed and model spectra respectively, and 111.4 is the $e$-folding time of $^{56}$Co in days. In Figure~\ref{fig:JK_model}, the observed and model spectra at three epochs are shown. The strength of the [\ion{O}{1}] doublet closely matches the 15 M$_\odot$ model at 308.02 d, whereas at 366.40 and 391.79 d the strength aligns better with the 12 M$_\odot$ model. This suggests that the progenitor mass of SN~2023zcu likely falls between 12 and 15 M$_\odot$.

\subsection{Bolometric Light-Curve Modeling}
\label{sec:bol_LC_modeling}

\begin{table}
	\begin{center}
	\caption{The best-fit core parameters with 2$\sigma$ uncertainties for the bolometric light-curve modeling of SN~2023zcu.
    }
        \label{tab:N & V results}
        \setlength{\tabcolsep}{3pt}

	\begin{tabularx}{\columnwidth}{lcc}
 
		\hline
		Parameter & Best-fit value & Prior range\\
            \hline
            \vspace{0.3cm}
            Initial radius [$R_0$ ($10^{13}$ cm)] & $3.86^{+0.14}_{-2.34}$ & 1--4 \\
            \vspace{0.2cm}
             Ejecta mass [$M_\mathrm{ej}$ (M$_\odot$)] & $10.01^{+0.34}_{-0.10}$ & 8--20\\
            \vspace{0.2cm}
            Kinetic energy [$E_{\rm k}$ (10$^{51}$ ergs)] & $1.94^{+0.06}_{-0.06}$ & 1.5--2.0 \\
            \vspace{0.2cm}
            Thermal energy [$E_\mathrm{th}$ (10$^{51}$ ergs)] & $0.36^{+0.53}_{-0.04}$ & 0.01--1\\
            \hline
		
	\end{tabularx}
    
        \end{center}
\end{table}

\begin{figure}
    \centering
    \includegraphics[width=\linewidth]{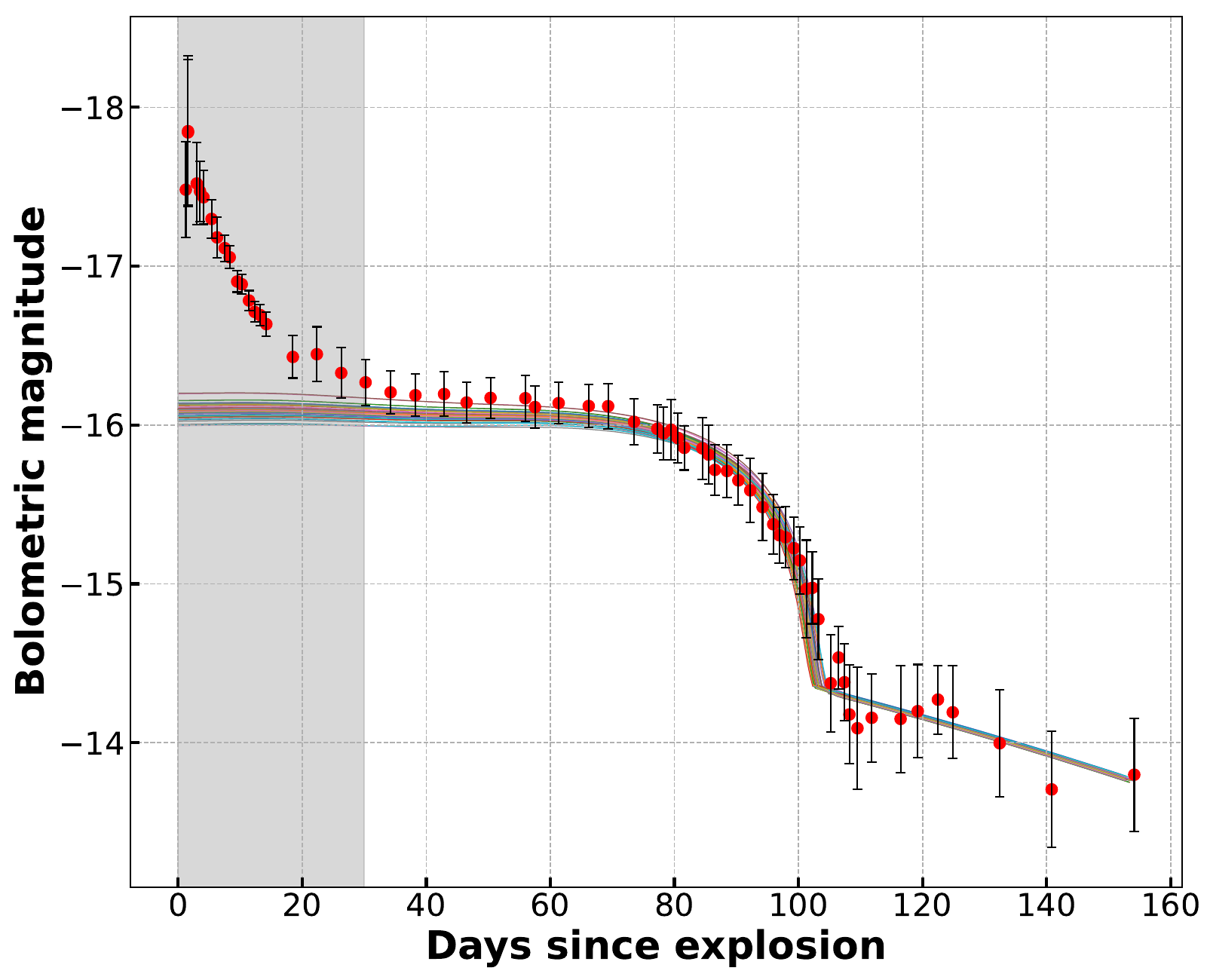}
    \caption{The 50 best-fit model light curves following \cite{Jager_2020} are shown along with the bolometric light curve of SN~2023zcu. The early phase (gray region) is excluded from the modeling as the shell component is not considered.}
    \label{fig:NandV}
\end{figure}

We employ a semi-analytical model to the bolometric light curve of SN~2023zcu in order to estimate the progenitor properties. The model provided by \cite{Nagy_2014} was further modified by \cite{Jager_2020} to include the MCMC routine in the numerical fitting code. In this model, a spherically symmetric SN expands homologously, with a uniform dense core and a shell with exponentially declining density. \cite{Jager_2020} only considers the core part, and hence the early light curve is not considered in the fit. During modeling, four main properties of the progenitor are sampled: radius ($R_0$), ejecta mass ($M_{\rm ej}$), kinetic energy ($E_{\rm k}$), and thermal energy ($E_{\rm th}$). The recombination temperature and Thomson scattering opacity ($\kappa$) are kept at 6000 K and 0.3 g cm$^{-2}$, respectively. As SN~2023zcu exhibits a normal plateau length, the effect of a magnetar is not considered while performing the modeling. The plateau phase of the synthetic light curve highly relies on the parameters $M_{\rm ej}$, $R_0$, and $E_{\rm k}$, while the tail part depends on the $^{56}$Ni mass and gamma-ray leakage. The parameters ($M_{\rm ej}$ - $E_{\rm k}$ and $E_{\rm th}$ - $R_0$) are well correlated based on the Pearson correlation coefficient. The 50 best-fit model light curves to the observed bolometric light curve of SN~2023zcu are displayed in Figure~\ref{fig:NandV}. The model considers the data beyond 30 d since the explosion. The plateau, drop from the plateau, and nebular phase are well reproduced by the model. This results in an ejecta mass of $\sim$10 M$_\odot$ for the progenitor, with an energy of $\sim 1.9 \times 10^{51}$ ergs and a $^{56}$Ni mass of $0.024 \pm 0.006$ M$_\odot$, consistent with the estimated $^{56}$Ni mass mentioned in Section~\ref{sec:Nimass}. The obtained values of the best-fit core parameters with 2$\sigma$ uncertainties and prior ranges are listed in Table~\ref{tab:N & V results}. The total energy obtained from semi-analytical modeling is higher than the typical energy observed in Type~IIP SNe. For example, the median value of energy is found to be $\sim$ 0.6$\times$10$^{51}$ ergs \citep{Martinez_2022} from hydrodynamical modeling. This can be plausibly due to the simplified assumptions in the semi-analytical model, which results in a discrepancy in the energy, especially when compared to that of hydrodynamical modeling estimations (SN~2017eaw; \citealt{Szalai_2019}, SN~2018gj; \citealt{Teja_2018gj}, SN~2018is; \citealt{raya_2018is}, SN~2018pq; \citealt{2018pq_2025}). 

\section{Conclusions}
\label{sec:discussion}

This paper presents high-cadence photometric and spectroscopic observations, along with an in-depth analysis, of a Type IIP SN~2023zcu. The observations span from the early to late nebular phase, allowing for a detailed study of the SN evolution and constraining the progenitor properties. SN~2023zcu falls into an intermediate-luminosity SN category with $M_{V50} = -16.13\pm0.02$ mag. It exhibits an optical thick phase of $\sim$100 d after the explosion and declines at a rate of $0.84 \pm 0.01$ mag (100 d)$^{-1}$ in the {\em V} band. The early light curve is modeled by the shock cooling model. The distance estimated from this model (26$^{+1}_{-2}$ Mpc) aligns well with the mean distance of the galaxy ($26.8 \pm 1.2$ Mpc in NED) as well as the tailored-EMP distance ($27.79 \pm 2.0$ Mpc). The progenitor radius results in $\sim$1200 R$_\odot$, which is higher than the radius derived from the semi-analytical modeling; hence, the shock cooling model estimated radius can be considered as an upper limit.

The rapid spectroscopic follow-up of the SN was started 1.19 d after the explosion. In the early spectra, a weak emission line on the top of the broad and extended H$\alpha$ profile is seen; however, the other high-ionization lines (e.g., \ion{He}{2}, \ion{C}{4}, \ion{C}{3}, \ion{N}{5}) are absent. An asymmetric `ledge-shaped' feature also appeared around 4500--4800 \AA\ during the early spectral evolution, similar to that found in SNe~2021yja and 2023axu. All these early spectral signatures suggest a low-level CSM interaction with the ejecta during the early-phase SN evolution. The early spectra were also compared with the model spectra by \cite{Dessart_2017} and found to closely resemble the model computed for a compact progenitor in a low-density CSM environment with (\texttt{r1w1h}) and without (\texttt{r1w1}) an extended atmosphere. This further strengthens the presence of low-density CSM. During the optical thick phase, a prominent H$\alpha$ P-Cygni profile is consistently visible throughout the spectral evolution, resembling the Type IIP SN characteristics. Metal lines (e.g., \ion{Fe}{2}, \ion{Ca}{2} NIR triplet) become more prominent as the SN evolves, and the photosphere recedes deeper into the ejecta. During the nebular phase, several forbidden lines ([\ion{Ca}{2}] 7291, 7324 \AA, [\ion{O}{1}] 6300, 6364 \AA, and weak \ion{Mg}{1}] 4571 \AA) are visible, as for the typical Type IIP SN~1999em. The physical parameters, such as temperature, velocity, and density profile in the early photospheric phase, are derived from the \textsc{TARDIS} modeling. \textsc{TARDIS} results confirm that SN~2023zcu is a normal Type IIP SN with a weak CSM interaction.

The metallicity of the SN has been determined by measuring the pEW of the \ion{Fe}{2} 5018 \AA\ line, which results in a solar metallicity, similar to the SN~1995ad \citep{Inserra_1995ad_2013}. SN~2023zcu exhibits velocity evolution similar to that of SNe~2006bp and 2021yja; however, it shows higher velocities compared to the Type IIP prototype SN~1999em. The velocity evolution during the first month is well aligned with the velocity determined with \textsc{TARDIS} modeling. The `Cachito' feature is visible during the plateau phase. This feature is produced by the interaction of ejecta with either the CSM or the RSG wind. The mass of the progenitor was constrained using two methods: comparing the nebular spectra with the models presented by \cite{Jerkstrand_2014}, and modeling the bolometric light curve. Nebular spectra comparison suggests that the progenitor mass of SN 2023zcu likely falls between 12 and 15 M$_\odot$. Semi-analytical modeling of the bolometric light curve gives an ejecta mass of approximately 10 M$_\odot$. Assuming a proto-neutron-star mass of about 2 M$_\odot$, this implies a progenitor mass of roughly 12 M$_\odot$. Hence, from both methods, the progenitor mass is consistent.

The overall analysis indicates that SN~2023zcu is a typical Type IIP SN. Due to the availability of high-cadence photometric and spectroscopic observations across all phases of its evolution, both the progenitor properties and explosion parameters are tightly constrained. Consequently, SN~2023zcu stands out as a well-studied Type IIP event and is a valuable addition to the existing Type IIP SNe sample.

\begin{acknowledgments}

The authors thank the reviewer for providing constructive comments on the manuscript, which have improved the clarity of presentation. This work uses data from the Las Cumbres Observatory global telescope network. The LCO group is supported by U.S. National Science Foundation (NSF) grants AST-2308113 and AST-1911151. M.D. acknowledges the Innovation in Science Pursuit for Inspired Research (INSPIRE) fellowship award (DST/INSPIRE Fellowship/2020/IF200251) for this work. K.M. acknowledges the support from the BRICS grant DST/ICD/BRICS/Call-5/CoNMuTraMO/2023 (G) funded by the Department of Science and Technology (DST), India. Time-domain research by the University of Arizona team and D.J.S. is supported by U.S. NSF grants 2108032, 2308181, 2407566, and 2432036, and by the Heising-Simons Foundation under grant \#2020-1864. N.F. acknowledges support from the U.S. NSF Graduate Research Fellowship Program under Grant DGE-2137419. K.A.B. is supported by an LSST-DA Catalyst Fellowship; this publication was thus made possible through the support of Grant 62192 from the John Templeton Foundation to LSST-DA.
A.V.F.’s research group at UC Berkeley acknowledges financial assistance from the Christopher R. Redlich Fund, as well as donations from Gary and Cynthia Bengier, Clark and Sharon Winslow, Alan Eustace and Kathy Kwan,   William Draper, Timothy and Melissa Draper, Briggs and Kathleen Wood, and Sanford Robertson (W.Z. is a Bengier-Winslow-Eustace Specialist in Astronomy, T.G.B. is a Draper-Wood-Robertson Specialist in Astronomy), and numerous other donors. A.V.F. is grateful for the hospitality of the Hagler Institute for Advanced Study as well as the Department of Physics and Astronomy at Texas A\&M University during part of this investigation.

Based partly on observations obtained as part of (GN-2024A-Q-403; P.I. Jennifer Andrews) at the international Gemini Observatory, a program of the U.S. NSF NOIRLab, which is managed by the Association of Universities for Research in Astronomy (AURA) under a cooperative agreement with the U.S. NSF on behalf of the Gemini Observatory partnership: the U.S. NSF (United States), National Research Council (Canada), Agencia Nacional de Investigaci\'{o}n y Desarrollo (Chile), Ministerio de Ciencia, Tecnolog\'{i}a e Innovaci\'{o}n (Argentina), Minist\'{e}rio da Ci\^{e}ncia, Tecnologia, Inova\c{c}\~{o}es e Comunica\c{c}\~{o}es (Brazil), and Korea Astronomy and Space Science Institute (Republic of Korea). These observations were processed using DRAGONS (Data Reduction for Astronomy from Gemini Observatory North and South). This work was enabled by observations made from the Gemini North telescope, located within the Maunakea Science Reserve and adjacent to the summit of Maunakea. We are grateful for the privilege of observing the Universe from a place that is unique in both its astronomical quality and its cultural significance. In addition, some of the data presented herein were obtained at Keck Observatory on Maunakea, which is a private 501(c)3 nonprofit organization operated as a scientific partnership among the California Institute of Technology, the University of California, and the National Aeronautics and Space Administration. The Observatory was made possible by the generous financial support of the W. M. Keck Foundation. 

Some observations reported here were obtained at the MMT Observatory, a joint facility of the University of Arizona and the Smithsonian Institution.

A major upgrade of the Kast spectrograph on the Shane 3 m telescope at Lick Observatory, led by Brad Holden, was made possible through generous gifts from the Heising-Simons Foundation, William and Marina Kast, and the University of California Observatories.
KAIT and its ongoing operation were made possible by donations from Sun Microsystems, Inc., the Hewlett-Packard Company, AutoScope Corporation, Lick Observatory, the U.S. NSF, the University of California, the Sylvia \& Jim Katzman Foundation, and the TABASGO Foundation. Research at Lick Observatory is partially supported by a generous gift from Google.  

\end{acknowledgments}

\software{
\texttt{NumPy} \citep{numpy}, \texttt{Astropy} \citep{astropy_2018},
\texttt{MatPLOTLIB} \citep{matplotlib_2007}, \texttt{IPython} \citep{ipython}, \texttt{emcee} \citep{emcee_Foreman},
\texttt{BANZAI} \citep{McCully_2018}, \texttt{floydsspec} pipeline \citep{Valenti_2014}, \texttt{alfoscgui} pipeline \citep{alfoscgui_2014}, \texttt{KastShiv} pipeline \citep{Shivvers_KastShiv6}, \texttt{PESSTO} pipeline \citep{pessto_smartt_2015}, \texttt{Goodman HTS} pipeline \citep{SOAR_GoodmanPipeline}, \texttt{IRAF} \citep{Tody1986, Tody1993}, \texttt{DRAGONS} pipeline \citep{Labrie_2019}, \texttt{PySALT} pipeline \citep{Crawford_2010}, \texttt{LPipe} pipeline \citep{Perley_LRIS}, \texttt{Light Curve Fitting} \citep{Griffin_2022}  
}

\bibliography{2023zcu_reference}{}
\bibliographystyle{aasjournalv7}

\onecolumngrid

\end{document}